\documentclass[aip,amsmath,amssymb,reprint,cha]{revtex4-2}

\usepackage{graphicx}
\usepackage{xcolor}
\usepackage[english]{babel}
\usepackage{amsmath}
\usepackage[colorlinks=true, allcolors=blue]{hyperref}

\begin{document}

\title{Antiperiodicity in the Duffing--Holmes oscillator: symmetry origin, parity selection, and transitions}

\author{Arturo C. Marti}
\email{marti@fisica.edu.uy}
\affiliation{Facultad de Ciencias, Universidad de la Rep\'ublica,
Igu\'a 4225, Montevideo 11400, Uruguay}

\author{E.~D.~Leonel}
\affiliation{Departamento de F\'isica, Universidade Estadual Paulista
             (UNESP), Av.\ 24A 1515, Rio Claro, S\~ao Paulo, Brazil}

\date{\today}

\begin{abstract}
We investigate the origin and distribution of antiperiodicity ---
oscillations satisfying $x(t+T)=-x(t)$ --- in the periodically driven
Duffing--Holmes oscillator, combining analytical arguments with extensive
numerical exploration. Antiperiodic orbits are precisely the
periodic orbits invariant under the half-period shift
symmetry $S:(x,\dot{x},t)\mapsto(-x,-\dot{x},\,t+T_d/2)$ of the equations
of motion, with $T_d$ the driving period.
We map the antiperiodic regions across the plane spanned
by the amplitude and the frequency of the forcing, together with the
periodic and chaotic domains and the potential wells visited by each
orbit. The invariance under $S$ imposes a parity selection
rule, verified without exception across our parameter sweeps:
antiperiodic orbits lock to the drive only at odd multiples of the
forcing period. Periodic orbits that lack the antisymmetry occur instead
as conjugate pairs related by $S$, each orbit being the point reflection
of its twin.
Following the approach of Swift and Wiesenfeld, who
established that in this class of systems a symmetric orbit cannot
undergo a direct period doubling, we quantitatively analyze the exit
from the antiperiodic regime and explicitly identify the two
codimension-one bifurcations through which it takes place: a
supercritical pitchfork, in which the antiperiodic orbit splits into two
conjugate, symmetry-broken orbits, and a saddle-node bifurcation with an
antiperiodic saddle, in which the antiperiodic orbit disappears with its
symmetry intact.
Antiperiodicity thus emerges as the orbit-level manifestation of a
discrete symmetry of the driven system.
\end{abstract}

\maketitle

\begin{quotation}
Symmetry and its spontaneous breaking are among the most unifying ideas
in physics, shaping phenomena from phase transitions to particle masses.
Nonlinear driven oscillators offer an unexpectedly simple stage on which
this interplay can be watched at work. When a particle in a symmetric
double-well potential is shaken periodically, its long-time response need
not simply follow the drive: it may repeat only after several forcing
cycles, become chaotic, or --- most strikingly --- reproduce its own
negative after a fixed delay, a behavior known as
antiperiodicity. Here we show that in the Duffing--Holmes oscillator,
arguably one of the simplest chaotic systems of this kind, antiperiodicity is not
a curiosity of particular waveforms but the direct expression of a
discrete symmetry of the equations of motion, which exchanges the two
wells while shifting time by half a forcing period. Every stable
oscillation must take a stand with respect to this symmetry: either it
respects it, and is then antiperiodic and constrained to repeat only
after an odd number of forcing cycles, or it breaks it, in
which case it is always accompanied by a mirror twin --- a coexisting
orbit that restores, jointly, the symmetry each one lost. Mapping where
these two possibilities occur across the parameters of the system reveals
how symmetry and chaos organize one of the most
paradigmatic nonlinear systems in physics.
 \end{quotation}

\section{Introduction}

The Duffing--Holmes oscillator (the terminology is not unique: the
double-well system considered here is often cited simply as the Duffing
oscillator, while the case of vanishing linear term is also known in the
literature as the Ueda
oscillator~\cite{ueda1979randomly,ueda1991survey,bonatto2008chaotic}) is among the simplest continuous-time
nonlinear systems capable of sustaining chaotic dynamics: a damped,
periodically driven nonlinear oscillator with one and a half degrees of
freedom (a three-dimensional phase space), it meets the minimal
requirements for the existence of chaos
\cite{ChaosKorsch,bonatto2008chaotic,holmes1979nonlinear,holmes1980phase}.
Despite this formal simplicity, its dynamics displays a remarkably rich
phenomenology, including periodic and chaotic motion, multiple routes to
chaos, multistability, and hysteresis
\cite{strogatz2018nonlinear,argyris2015exploration}.

Among the regularities exhibited by this class of systems, a
particularly elusive one is the phenomenon of \textit{antiperiodic
oscillations}  \cite{freire2013antiperiodic,singla2015antiperiodic,long2016new,haraux1989antiperiodic}.
A function $x(t)$ is said to be antiperiodic if it
satisfies
\begin{equation}
    x(t + T) = -x(t) \quad \text{for all } t,
    \label{eq:antiperiodic}
\end{equation}
where $T$ is the antiperiod; an antiperiodic function is therefore
periodic with period $2T$, but carries an additional sign-reversal
structure. Trivial examples are the trigonometric functions $\sin t$
and $\cos t$. Nontrivial antiperiodic solutions, however, arise in
physically relevant nonlinear systems and exhibit a richness that is
far from obvious. In harmonically driven systems, antiperiodicity is intimately
connected with the invariance of the equations of motion under the
symmetry $S:(x,\dot{x},t)\mapsto(-x,-\dot{x},t+T_d/2)$, where
$T_d = 2\pi/\omega$ denotes the period of the drive, which maps the
phase space onto itself while shifting time by half a forcing period: antiperiodic orbits are precisely the periodic orbits
that are individually invariant under
$S$~\cite{sato1983universal,swift1984suppression,novak1982transition,dhumieres1982chaotic}.

In systems with this symmetry, periodic orbits are of
two kinds. Besides the antiperiodic orbits, which are individually
invariant under $S$, there are periodic orbits that lack the symmetry;
these cannot occur in isolation, since the image under $S$ of such an
orbit is a distinct orbit of the same period, so that symmetry-broken
periodic orbits always appear in conjugate pairs $A$, $B=S(A)$, each
being the point reflection of its twin shifted by half a forcing
period, and the pair jointly restores the symmetry that each member
lost. This dichotomy between orbits individually invariant
under $S$ and conjugate pairs related by it is not new. It was already
exploited by Sato, Sano, and Sawada~\cite{sato1983universal} to classify
periodic orbits of the single-well cubic Duffing oscillator under the
same transformation (up to the trivial rescaling $T_d/2\to\pi$ at
$\omega=1$); in their terminology the $S$-invariant orbits are called
\emph{symmetric} and the conjugate pairs \emph{antisymmetric}, the
reverse of the convention adopted here, where the $S$-invariant orbits
are the antiperiodic ones. A complementary spectral signature of the
same dichotomy was reported by Novak and
Frehlich~\cite{novak1982transition} in an analog-circuit Duffing
oscillator: the small-amplitude response, containing odd harmonics
only, is $S$-invariant by construction and loses stability, as the
drive amplitude increases, to a symmetry-broken state containing even
harmonics. The same spectral signature, odd multiples of the fundamental
frequency only, with even harmonics appearing when the symmetry is broken,
was identified in general by Swift and
Wiesenfeld~\cite{swift1984suppression}. An essentially identical mechanism, formalized through a Floquet stability analysis reducing to Hill's equation, was derived by D'Humi\`eres \emph{et al.}~\cite{dhumieres1982chaotic} for the forced
pendulum, where the same symmetry was shown analytically to break necessarily before any period-doubling bifurcation can occur.

In their pioneering paper, Swift and
Wiesenfeld~\cite{swift1984suppression} placed this phenomenology on a
rigorous, model-independent basis, proving that the driven damped
oscillator in any symmetric potential (the class to which
Eq.~\eqref{eq:duffing} belongs) cannot undergo a direct period-doubling
bifurcation from a symmetric orbit; we return to this general result
and to its consequences in
Secs.~\ref{sec:existence}--\ref{sec:fold}. None of these
studies, however, addresses a double-well potential, where the relation
of an orbit to $S$ is tied to which well or wells it visits; extending
the analysis to that case is one of the central themes developed
below.

A rigorous existence theory for antiperiodic solutions
of this symmetric class was developed by
Haraux~\cite{haraux1989antiperiodic} for broad families of odd,
non-monotone evolution equations forced by an antiperiodic term,
including first- and second-order systems in $\mathbb{R}^N$ and
semilinear wave equations. There, the antiperiodicity of a driven
solution follows almost for free from the symmetry of the equation
rather than from restrictive sign or growth conditions on the
nonlinearity, mirroring the role played by $S$ here. Crucially, Haraux
also shows that existence does not imply uniqueness once the underlying
potential is non-convex: for the scalar equation
$u'+u^3-u=\varepsilon\sin t$, whose odd part derives from the same
double-well potential considered below, he constructs two coexisting
solutions confined near each well that are exchanged by the symmetry
transformation, alongside the guaranteed antiperiodic one: an early,
rigorous instance, in a first-order system, of precisely the
antiperiodic/conjugate-pair dichotomy at the center of this work.

The organization of periodic stability domains into an
infinite hierarchy of nested spirals converging toward a periodicity
hub was first reported by Bonatto and
Gallas~\cite{bonatto2008periodicity} for a symmetric autonomous
electronic circuit; that work also documented the increasingly complex
periodic waveforms found along the spirals and explicitly distinguished
self-symmetric periodic orbits from symmetry-related conjugate pairs.
In our own earlier
work~\cite{cabeza2013periodicity,freire2013antiperiodic,freire2014self},
we investigated antiperiodic behavior in a modified Chua circuit with
piecewise-linear resistances, finding that antiperiodic oscillations
are a robust experimental phenomenon and not a numerical artifact. A
key result was that the antiperiodic solutions
of that circuit do not appear in isolation
but form an infinite family of waveforms of increasing complexity,
organized into the same kind of extended spiral-shaped
stability regions in parameter
space, all converging toward a single focal point that acts as an
organizing center for the entire hierarchy. Analysis of the maximum
Lyapunov exponents confirmed this picture~\cite{freire2014self}:
although the detailed structure of the diagrams depends on the
observable chosen, the global spiral topology is preserved, offering a
consistent view of the coexistence of ordered and chaotic dynamics.

Interest in antiperiodic oscillations, moreover, extends well beyond
purely academic motivations. In applied mathematics and engineering,
the existence and stability of antiperiodic solutions is an active
research topic in models of artificial neural networks --- including
shunting inhibitory cellular neural networks with oscillating leakage
coefficients~\cite{long2016new} and fuzzy delayed cellular
neural networks with impulses~\cite{Xu2020} --- where
antiperiodic regimes arise naturally from the signal transmission
between neurons and their stability properties bear directly on the
performance of the network. Antiperiodic structures are equally
familiar in physics and mathematics: fermionic fields at finite
temperature obey antiperiodic boundary conditions in imaginary time, a
cornerstone of the Matsubara formalism~\cite{kapusta2006finite}, while
antiperiodic boundary-value problems constitute an established line of
research in nonlinear analysis since the pioneering work of
Okochi~\cite{okochi1988existence}. A detailed understanding of how
antiperiodicity emerges, persists, and disappears in a simple
nonlinear oscillator is therefore of interest well outside nonlinear
dynamics itself.

The first experimental observation of antiperiodic oscillations
in a forced Duffing oscillator was reported by Shaw
\emph{et al.}~\cite{shaw2015antiperiodic}. In that work, the authors
combined numerical simulations, theoretical analysis, and analog
electronic circuit experiments to demonstrate the existence of robust
and tunable antiperiodic oscillations in a Duffing system with negative linear stiffness. A
key result was the identification of a peak-adding cascade, in which
the number of peaks in the antiperiodic oscillations systematically
increased, from three to twenty-five, as the forcing strength was
varied. To characterize these dynamics, bifurcation diagrams were
constructed using the forcing amplitude as the control parameter,
complemented by stability analyses of the associated fixed points.
Antiperiodicity was further verified through a similarity function,
while additional insight into the dynamics was obtained from
phase-space representations, Fourier spectra, and Hurst exponent
calculations based on rescaled range analysis.
These observations establish antiperiodicity as a genuine physical
regime; what remains open is \emph{why} and \emph{where} it must occur.

The present work centers on the existence of antiperiodicity in
the Duffing oscillator and on its distribution in parameter space, and
on how the antiperiodic regimes are organized together with the periodic
and chaotic ones.
That antiperiodic oscillations have already been observed
experimentally in the Duffing oscillator underscores the physical
relevance of these questions ---
such regimes are robust phenomena and not numerical artifacts ---
although the analysis developed here is structural and independent of
any particular realization. Specifically, we address three fundamental
questions:
(i) How does
the symmetry of the Duffing oscillator constrain its periodic
solutions, and what distinguishes the antiperiodic orbits from the
remaining periodic orbits? (ii) How
are antiperiodic regions distributed in parameter space and with
respect to the potential wells, and what regularities do they display? (iii) How does
antiperiodicity come to an end, i.e.\ through which bifurcations can an
antiperiodic orbit lose its stability: by shedding its symmetry and
giving way to a pair of conjugate, symmetry-broken solutions, or by
disappearing with its symmetry intact?

\section{Periodic and chaotic dynamics across the forcing parameter plane}
\label{sec:dynamics}

The Duffing--Holmes oscillator is described by
\begin{equation}
    \ddot{x} + \delta \, \dot{x} + \alpha\,x + \beta \,x^{3}
    = \gamma \cos(\omega t),
    \label{eq:duffing}
\end{equation}
where $\delta \geq 0$ is the damping coefficient, $\alpha$ governs the linear stiffness (negative values yield a double-well potential), and $\beta$ controls the degree of nonlinearity in the restoring force; in particular, for $\beta = 0$ and $\alpha>0$ Eq.~\eqref{eq:duffing} reduces to the damped driven harmonic oscillator. On the right-hand side, $\gamma$ is the amplitude of the periodic driving force (with $\gamma = 0$ corresponding to the unforced case) and $\omega$ is its angular frequency.

The conservative part of Eq.~\eqref{eq:duffing} derives from the potential
\begin{equation}
    V(x) = \tfrac{1}{2}\alpha x^{2} + \tfrac{1}{4}\beta x^{4},
    \label{eq:potential}
\end{equation}
which for $\alpha < 0$ and $\beta > 0$ describes a symmetric double well with two minima at $x = \pm\sqrt{-\alpha/\beta}$ separated by a local maximum at $x = 0$. Throughout this work we fix $\alpha = -1$, $\beta = 1$, and $\delta = 0.3$, so that the unforced system possesses two symmetric potential wells; the line $x = 0$ separating them will be used below to classify trajectories according to whether they remain confined to a single well or visit both. These parameter values are common to all the results and figures presented in this work and are therefore not repeated in the captions; the same applies to the numerical settings (integration scheme, tolerances, transient, and integration window), which are stated once in Appendix~\ref{app:numerical}; the captions quote only the settings specific to each figure (parameter ranges, grids, and sweep steps).

\begin{figure}[htbp]
\centering
\includegraphics[width=\columnwidth]{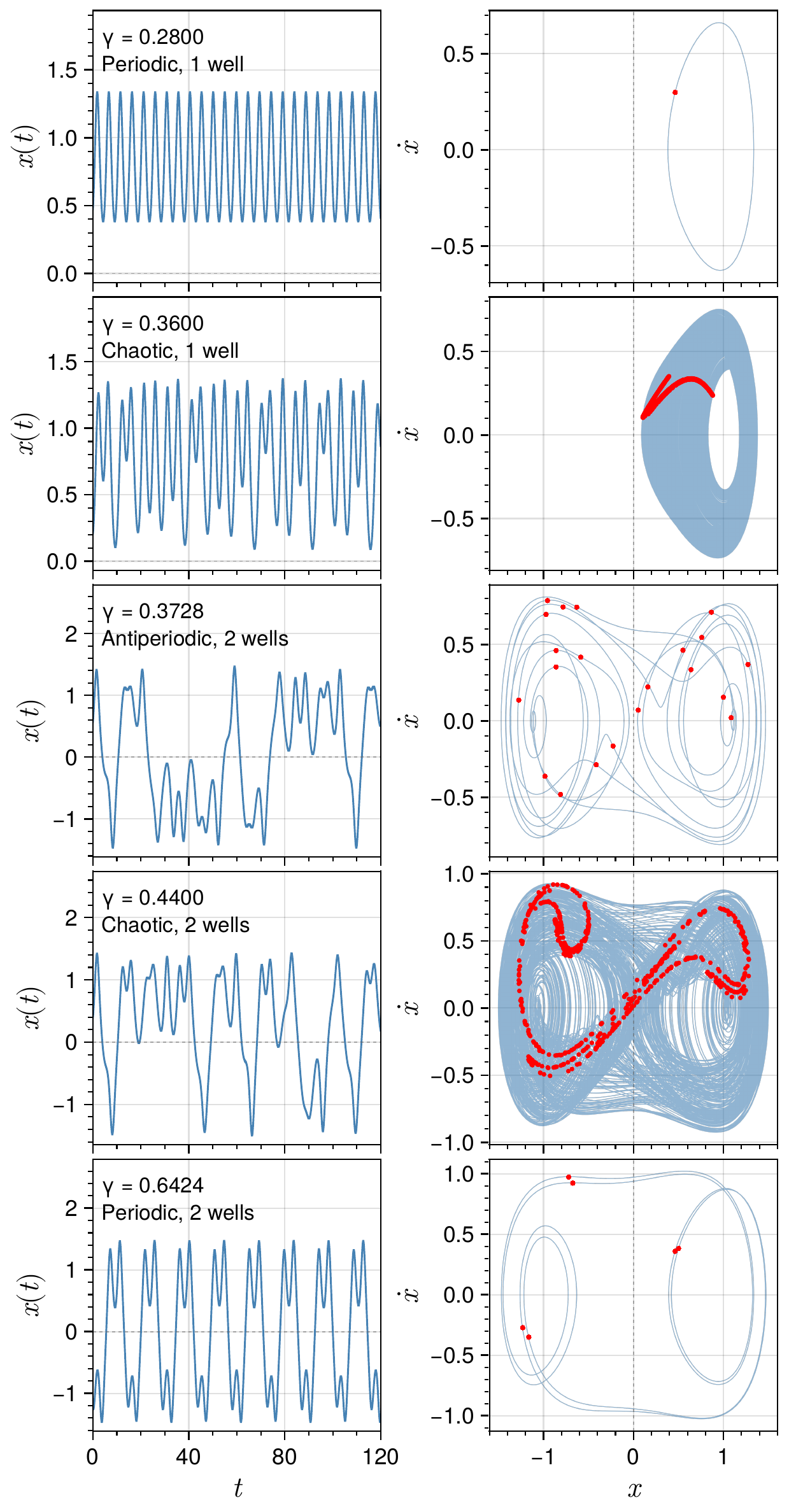}
\caption{Time series $x(t)$ (left panels) and phase portraits
$\dot{x}$ vs.\ $x$ (right panels) of the Duffing--Holmes
    oscillator for $\omega = 1.3$ and five representative values
    of the forcing amplitude $\gamma$.
    red dots in the phase portraits indicate the stroboscopic
    Poincar\'{e} map sampled at integer multiples of the drive
    period $T_d = 2\pi/\omega$.    From top to bottom:
    (\textit{i})~$\gamma = 0.2800$, periodic orbit confined to a
    single potential well (single Poincar\'{e} point);
    (\textit{ii})~$\gamma = 0.3600$, chaotic orbit confined to a
    single potential well (densely sampled Poincar\'{e} section
    within that well);
    (\textit{iii})~$\gamma = 0.3728$, antiperiodic orbit alternating
    between both potential wells, whose
    Poincar\'{e} section comprises 21 points spread across both
    wells, i.e.\ an antiperiodic orbit of order $k=21$;
    (\textit{iv})~$\gamma = 0.4400$, chaotic orbit exploring both
    potential wells (densely sampled Poincar\'{e} section spanning
    $x < 0$ and $x > 0$);
    (\textit{v})~$\gamma = 0.6424$, periodic orbit visiting both
    wells, with a sparse, low-order Poincar\'{e} section (three
    points) spanning both wells.
}
    \label{fig:5casos}
\end{figure}

To illustrate the variety of dynamical regimes exhibited by Eq.~\eqref{eq:duffing}, Fig.~\ref{fig:5casos} shows time series and phase portraits for five representative values of the forcing amplitude $\gamma$ at fixed $\omega = 1.3$, spanning periodic and chaotic motion confined to a single well, periodic and chaotic motion visiting both wells, and an antiperiodic orbit.

In the periodic panels, (\textit{i}) confined to a single well and
(\textit{v}) visiting both wells, the trajectory settles onto a closed curve, and the
number of stroboscopic points equals the ratio
\begin{equation}
  k=\frac{T_{\mathrm{orb}}}{T_d}
  \label{eq:order}
\end{equation}
between the period
of the orbit and the driving period, which we call the \emph{order} of
the orbit. The antiperiodic
case, (\textit{iii}), displays the characteristic half-period anti-symmetry: in the time series
the signal reproduces its own negative after half a period,
$x(t+T_{\mathrm{orb}}/2)=-x(t)$, while in the phase portrait this becomes
invariance under the inversion $(x,\dot{x})\to(-x,-\dot{x})$. Its
order $k=21$ means that the orbit locks to the
drive with $T_{\mathrm{orb}}=21\,T_d$; moreover, each maximum of $x(t)$
corresponds to a downward crossing of the $\dot{x}=0$ axis in the phase portrait,
and the 17 maxima counted within one orbital period coincide with the 17
such crossings. The odd count is not accidental: the half-period shift
maps maxima onto minima and the two alternate along the orbit, so that an
antiperiodic orbit with nondegenerate extrema necessarily displays an odd
number of maxima per period, consistent with the odd peak numbers of the
peak-adding cascade of Ref.~\cite{shaw2015antiperiodic}.

The chaotic panels, (\textit{ii}) and (\textit{iv}), in contrast, show a
time series that never repeats and stroboscopic points that never revisit the
same location: for arbitrarily long integration they would progressively fill a
bounded region of the section, the cross-section of the underlying strange
attractor. Two qualitatively different chaotic regimes are apparent: in one,
(\textit{ii}), the
motion remains confined to a single well, whereas in the other,
(\textit{iv}), it intermittently
crosses the potential barrier, hopping between the two wells.
Cases (\textit{iv}) and (\textit{v}) lie just beyond the two exits of
antiperiodic windows analyzed in Secs.~\ref{sec:fold}
and~\ref{sec:transition}, respectively: the chaotic attractor that follows
the saddle-node termination of the window of order $k=5$, and the
symmetry-broken orbit of order $k=3$ created in the pitchfork that ends
the window of order $k=3$.

\begin{figure}[htbp]
    \centering
\includegraphics[width=\columnwidth]{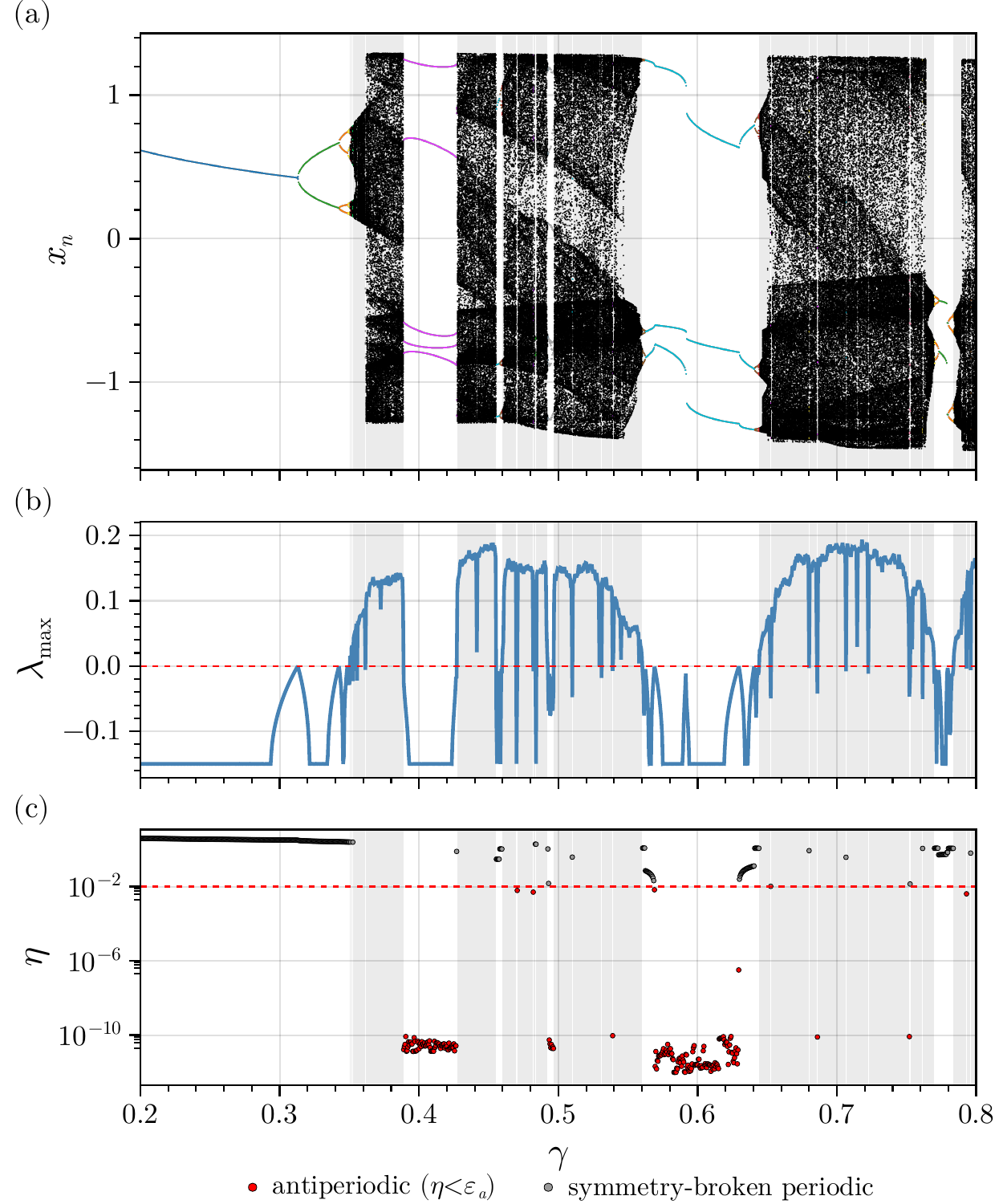}
\caption{ \label{fig:bifurcacion_omega13}
Bifurcation diagram (a) and maximum Lyapunov exponent $\lambda_{\max}$ (b)
as a function of the  forcing amplitude $\gamma$, for fixed $\omega = 1.3$,    swept with step $\Delta\gamma=5\times10^{-4}$.
    Panel~(a) shows the stroboscopic section $x_n$, recorded after
    the transient while the attractor is followed by continuation across
    successive values of $\gamma$.
    Colors of the periodic branches encode the orbital order $k$
    resolved in the stroboscopic section: period-1 (blue),
    period-2 (green), period-3 (cyan), period-4 (orange), and
    higher orders in additional distinct colors. Chaotic motion is
    shown in black and is defined by a positive maximum Lyapunov
    exponent, consistently with panel~(b), where the dashed
    red line marks $\lambda_{\max} = 0$.
    Panel~(c) displays the antiperiodicity residual
    $\eta$ [Eq.~\eqref{eq:antip_residual}], computed for the periodic
    orbits only, on a logarithmic scale: red circles indicate
    antiperiodic orbits ($\eta<\varepsilon_{a}$) and gray circles
    symmetry-broken periodic orbits, the dashed red line marking the
    threshold $\varepsilon_{a}$. Chaotic windows, where $\eta$
    is undefined, are shaded as vertical bands across the three
    panels.
  }
\end{figure}

\begin{figure}[htbp]
    \centering
\includegraphics[width=\columnwidth]{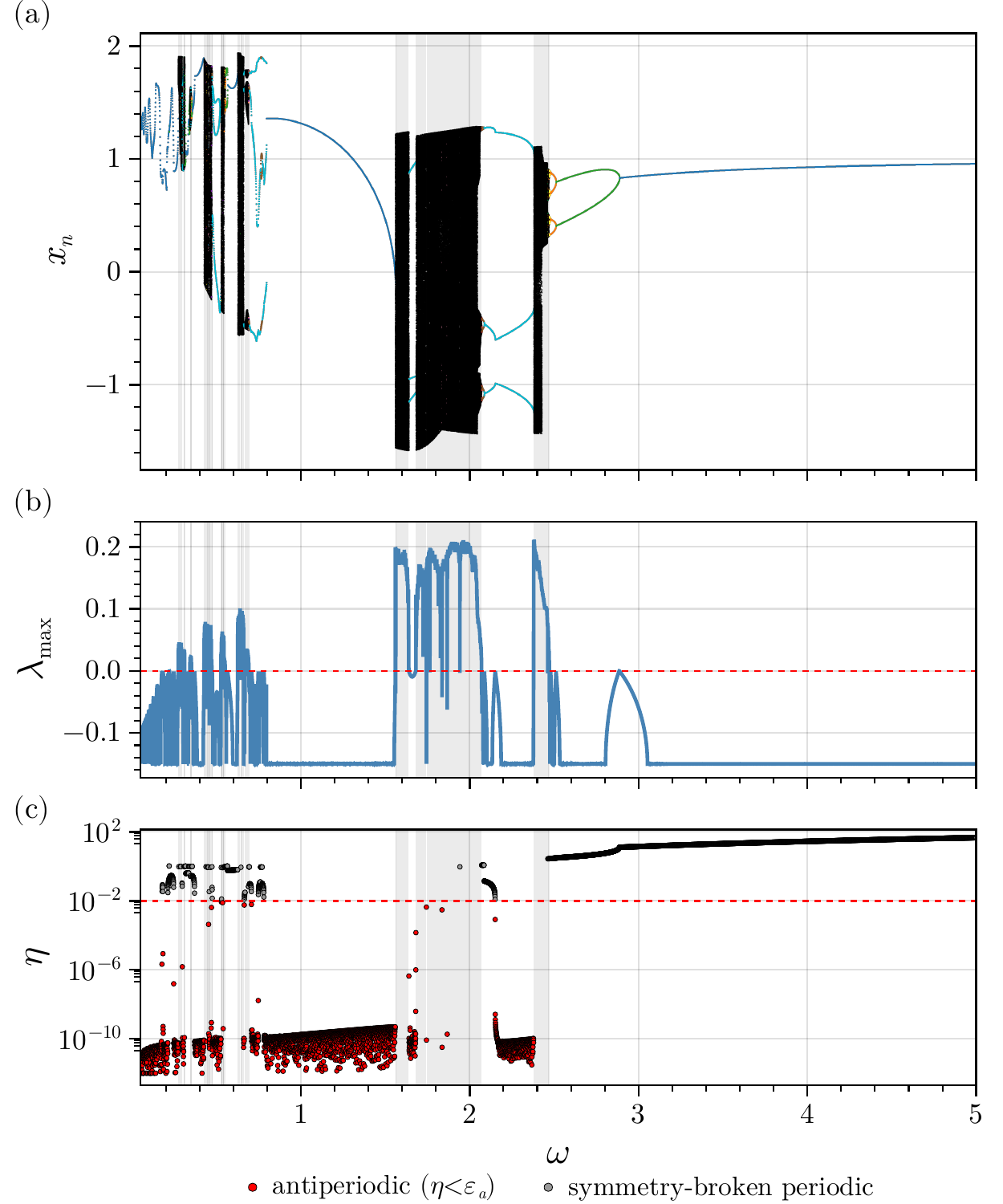}
\caption{
  \label{fig:bifurcacion_gamma1}
  Same as Fig.~\ref{fig:bifurcacion_omega13}, but as a
    function of the forcing frequency $\omega$, for fixed
    $\gamma = 1.0$: bifurcation diagram [panel~(a)], maximum Lyapunov
    exponent $\lambda_{\max}$ [panel~(b)], and antiperiodicity residual
    $\eta$ [panel~(c)], with the same construction, color coding, and
    thresholds in all panels.
    The frequency is swept over
    $\omega \in [0.05,\,5.0]$, with step
    $\Delta\omega=5\times10^{-4}$, the lower bound avoiding the
    singularity of the drive period at $\omega = 0$.}
\end{figure}

The organization of these regimes across parameter space is summarized by
two bifurcation diagrams, constructed by recording the values $x_n$ of the
stroboscopic Poincar\'{e} section defined by the forcing,
$x_n = x(nT_d)$, after the transient has been discarded.
Figure~\ref{fig:bifurcacion_omega13} shows the diagram obtained by varying
the forcing amplitude $\gamma$ at fixed $\omega = 1.3$, while
Fig.~\ref{fig:bifurcacion_gamma1} shows the diagram obtained by varying
the forcing frequency $\omega$ at fixed $\gamma = 1.0$. In both figures
panel~(b) displays the corresponding maximum Lyapunov exponent
$\lambda_{\max}$, which identifies the windows of chaotic motion,
and panel~(c) displays, on a logarithmic scale, the antiperiodicity
residual
\begin{equation}
  \eta=\frac{1}{A}\,
  \Big\langle\,\big|\,x(t)+x\!\left(t+\tfrac{1}{2}T_{\mathrm{orb}}\right)\big|\,\Big\rangle ,
  \label{eq:antip_residual}
\end{equation}
where $A$ is half the peak-to-peak excursion of $x$ and the average runs
over the last three orbital periods of the recorded window; $\eta$ vanishes
for an antiperiodic orbit and is of order unity for a periodic
orbit that breaks the symmetry. This third panel
resolves, within the periodic windows, the two possible responses of the
orbit to the symmetry $S$, antiperiodic ($\eta<\varepsilon_{a}$, where
$\varepsilon_{a}$ is the antiperiodicity threshold defined in
Appendix~\ref{subsec:descriptors}) or
symmetry-broken; $\eta$ is undefined in the chaotic windows, which are
shaded across the three panels. The
diagrams are computed following the attractor in the direction of
increasing control parameter: the final state reached at each parameter
value seeds the integration at the next one, so that a single attractor
branch is tracked by numerical continuation (Appendix~\ref{subsec:scan}).


Both diagrams display an alternation of chaotic and regular intervals that
form a complex, finely interleaved structure. Manifestations of
multistability are also visible: branches appear or disappear abruptly as
the control parameter is varied, signalling that the trajectory has jumped
from one attractor to a coexisting one --- the multistability-induced
discontinuities characteristic of continuation sweeps in this system.
Notably, these jumps are not a numerical artifact: the same abrupt
transitions between coexisting attractors have been observed
experimentally in an analog electronic implementation of the
Duffing--Holmes oscillator~\cite{gargiulo2026order}. Superimposed on this
structure, period-doubling and period-adding cascades are observed,
together with windows of periodicity embedded in the chaotic intervals.
Throughout both diagrams the classification is coherent with
the Lyapunov panels~(b): regular intervals correspond to $\lambda_{\max} < 0$ and chaotic
intervals to $\lambda_{\max} > 0$, confirming that the color coding of
panels~(a) and the Lyapunov diagnostics identify the same transitions.

The frequency diagram of Fig.~\ref{fig:bifurcacion_gamma1} deserves a
separate comment. For $0.27\lesssim\omega\lesssim0.54$ the diagram displays
a dense alternation of chaotic and antiperiodic intervals, many of the
latter narrower than the sweep step; over the whole sweep the antiperiodic
points form $26$ intervals with $51$ boundaries, $37$ of them with chaotic
motion and $13$ with a periodic neighbor. In every one of the latter the
neighbor is a symmetry-broken orbit of the same order $k$, with
$\lambda_{\max}$ close to zero and $\eta$ just above $\varepsilon_{a}$ at
the boundary, features
consistent with the pitchfork analyzed in
Sec.~\ref{sec:transition}. Depending on the direction in which the
bifurcation is traversed, the antiperiodic orbit either loses its
symmetry, as at $\omega\approx0.18$ and $0.21$ ($k=1$) and at
$\omega\approx0.67$ and $0.75$ ($k=3$), or recovers it, as at
$\omega\approx2.15$ ($k=3$), where the restoration is the last step of
an inverse period-doubling cascade that unfolds entirely on the
symmetry-broken branch, consistently with the suppression of period
doubling for $S$-invariant orbits. The remaining exits lead to chaotic
motion, as at $\omega\approx1.56$ and $2.38$,
as found beyond the saddle-node termination analyzed in
Sec.~\ref{sec:fold}, although the sweep alone does not identify the
bifurcation responsible for each of these boundaries. In no case
is an antiperiodic orbit adjacent to a periodic orbit of order $2k$.

\begin{figure}[htbp]
    \centering
\includegraphics[width=0.49\textwidth]{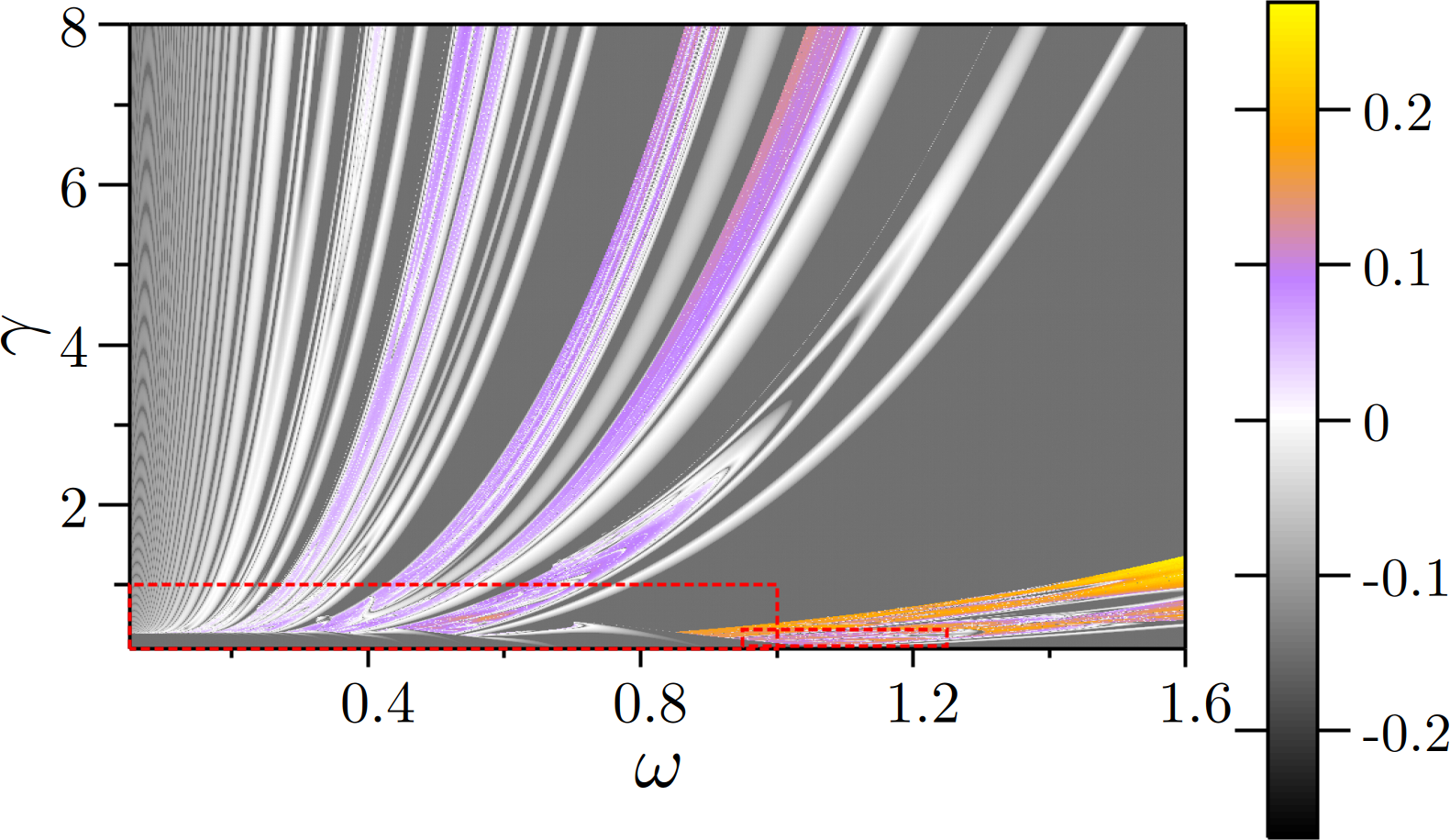}
\caption{
\label{fig:global}
Maximum Lyapunov exponent in the forcing parameter plane $(\omega, \gamma)$,
computed on an $800\times800$ grid over $\omega\in[0.05,1.6]$ and
$\gamma\in[0.2,8.0]$.
Black and gray tones indicate periodic  motion and purple--orange--yellow
tones indicate chaotic dynamics. The boxes outlined with dashed red lines indicate
the regions considered in Figs.~\ref{fig:shrimps_maps}--\ref{fig:origen_regime}.
}
\end{figure}

A more complete picture of how these regimes are organized is obtained by
scanning simultaneously the forcing amplitude $\gamma$ and the driving
frequency $\omega$. Figure~\ref{fig:global} shows the maximum Lyapunov
exponent over a broad region of the $(\omega,\gamma)$
plane, providing an overall map of
the regular and chaotic domains discussed above: chaos concentrates in
well-defined regions immersed in a periodic background, with intricate
boundaries between the two. Guided by this global view, two regions are
selected for closer inspection in the following figures. The first
contains shrimp-shaped periodic structures embedded in the chaotic
domain, of the kind identified in this system by Bonatto
\emph{et al.}~\cite{bonatto2008chaotic}; the second is the region close to the
origin (small values of $\omega$), where the driving period becomes long
compared with the intrinsic time scale of the oscillator and the
parameter plane acquires a particularly fine structure. For each region
three complementary maps are presented: the maximum Lyapunov exponent
$\lambda_{\max}$, the order $k$ of
the periodic orbits, and the number of peaks per period.

The robustness of these maps with respect to the sweep strategy
(attractor tracking along either axis, or fixed initial conditions) is
assessed in Appendix~\ref{subsec:robustness}: the regime classification
is essentially insensitive to it, whereas the well selected by
single-well orbits is not, a direct signature of multistability.

\begin{figure}[htbp]
\centering
\includegraphics[width=0.45\textwidth]{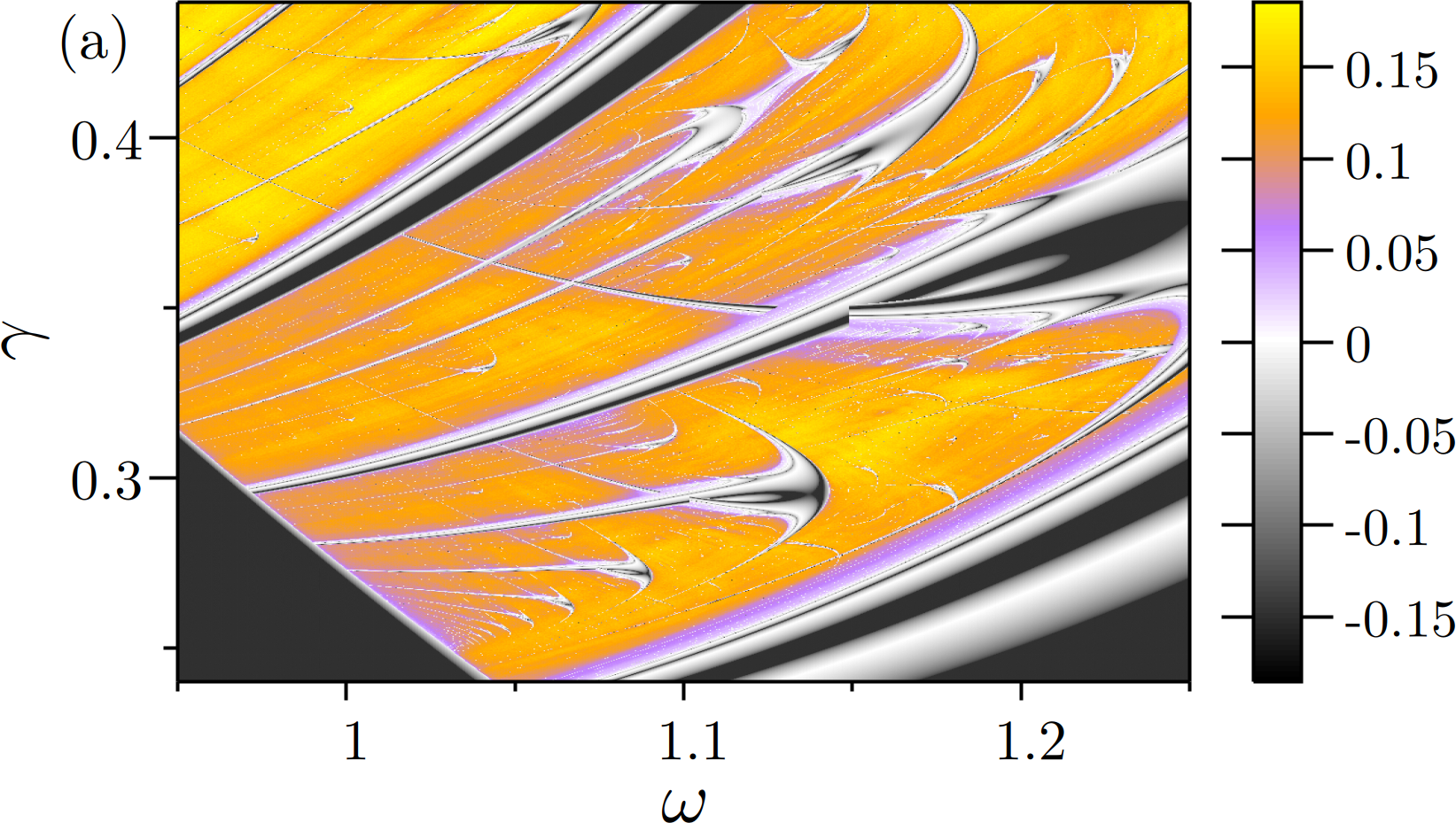}
\includegraphics[width=0.45\textwidth]{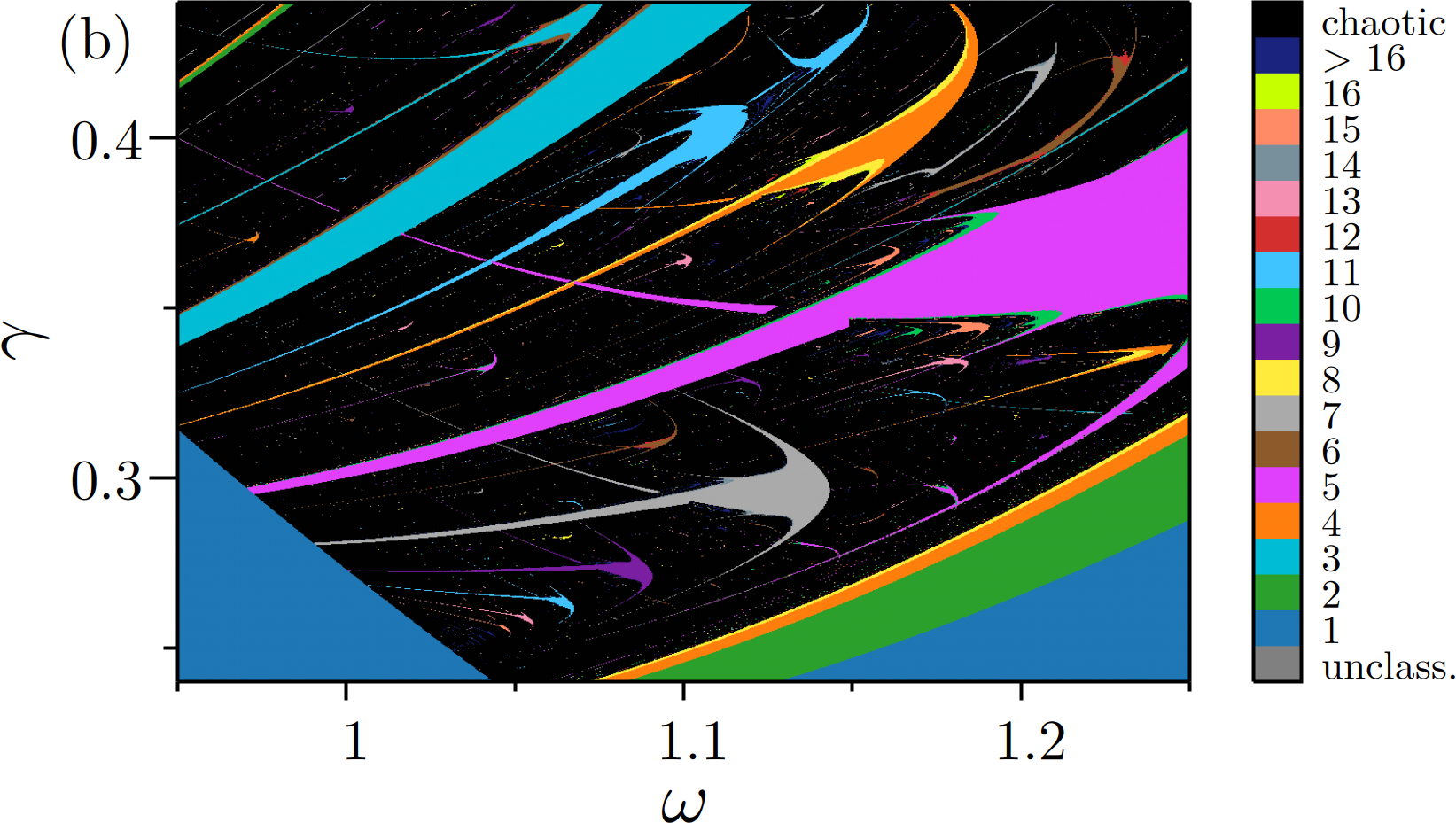}
\includegraphics[width=0.45\textwidth]{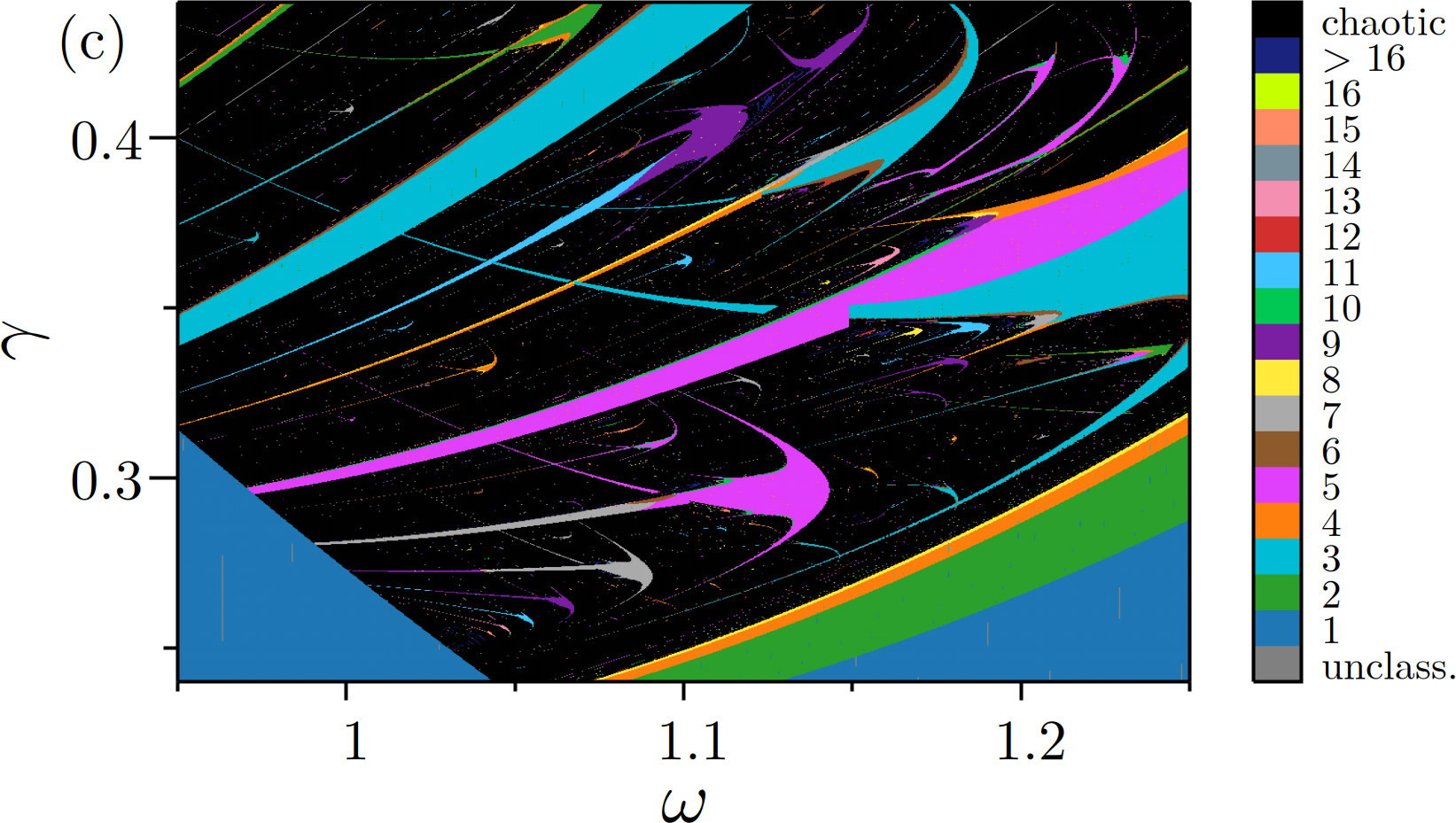}
\caption{
\label{fig:shrimps_maps}
Three complementary dynamical characterizations in the forcing parameter plane
$(\omega, \gamma)$, for one of the two regions marked with dashed boxes
in Fig.~\ref{fig:global}: the region containing the shrimp-shaped
periodic structures.
(a)~Maximum Lyapunov exponent $\lambda_{\max}$.
(b)~Orbit order $k$ [Eq.~\eqref{eq:order}], shown with a categorical
color scale for $k=1$ to $16$; orbits with $k>16$ share a single
dark-blue level, and chaotic regions are shown in black.
(c)~Number of peaks per period, with the same color scale as in~(b).
In panels (b) and (c), dark-gray dots mark unclassified points, i.e.\
non-chaotic points ($\lambda_{\max}\le\varepsilon_{\lambda}$) for which no
stroboscopic recurrence was detected within the recorded window
(Appendix~\ref{subsec:strobo}).
All panels: $800\times800$ grid over $\omega\in[0.95,1.25]$ and
$\gamma\in[0.24,0.44]$.}
\end{figure}

\begin{figure}[htbp]
\centering
\includegraphics[width=0.45\textwidth]{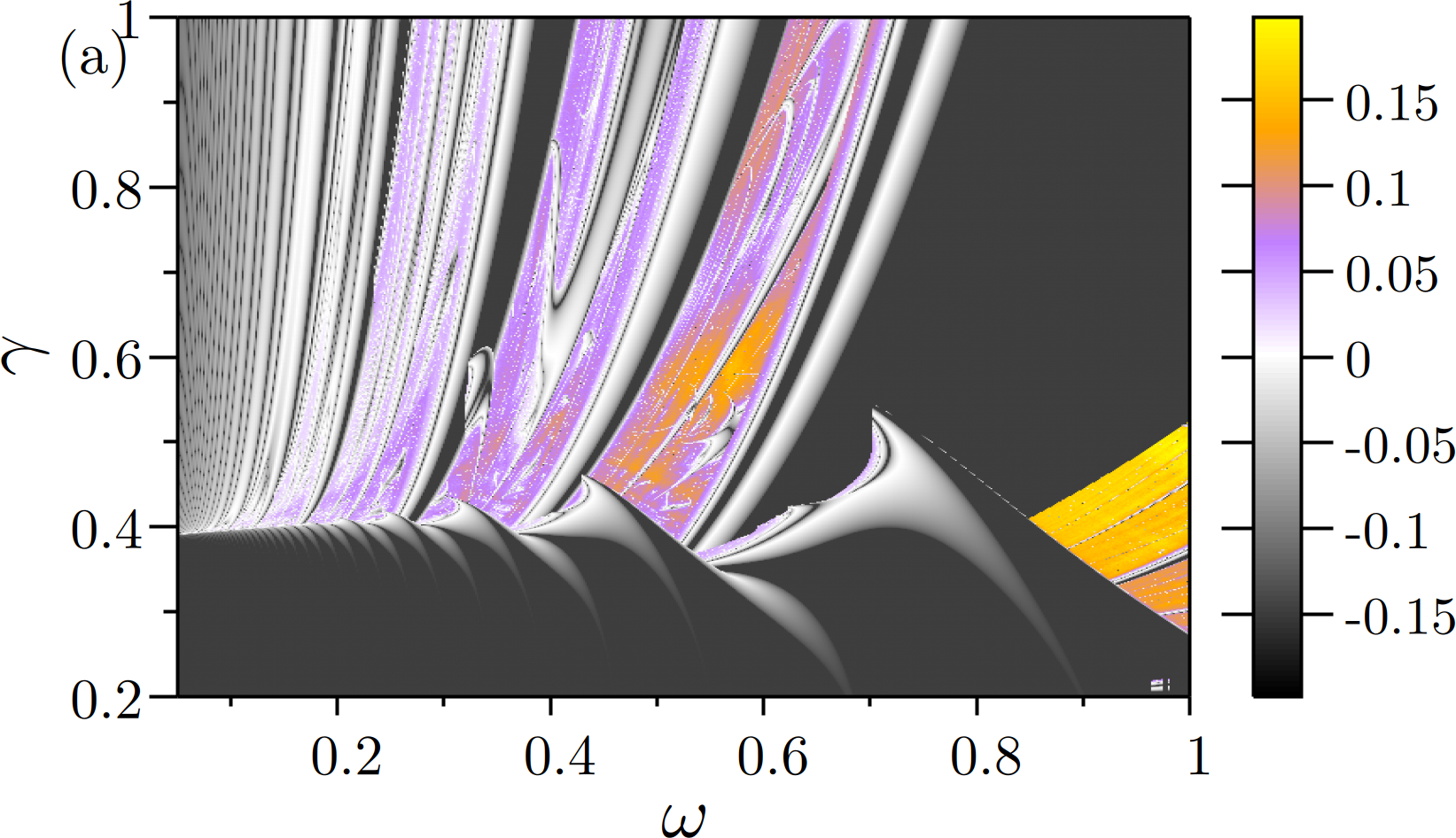}
\includegraphics[width=0.45\textwidth]{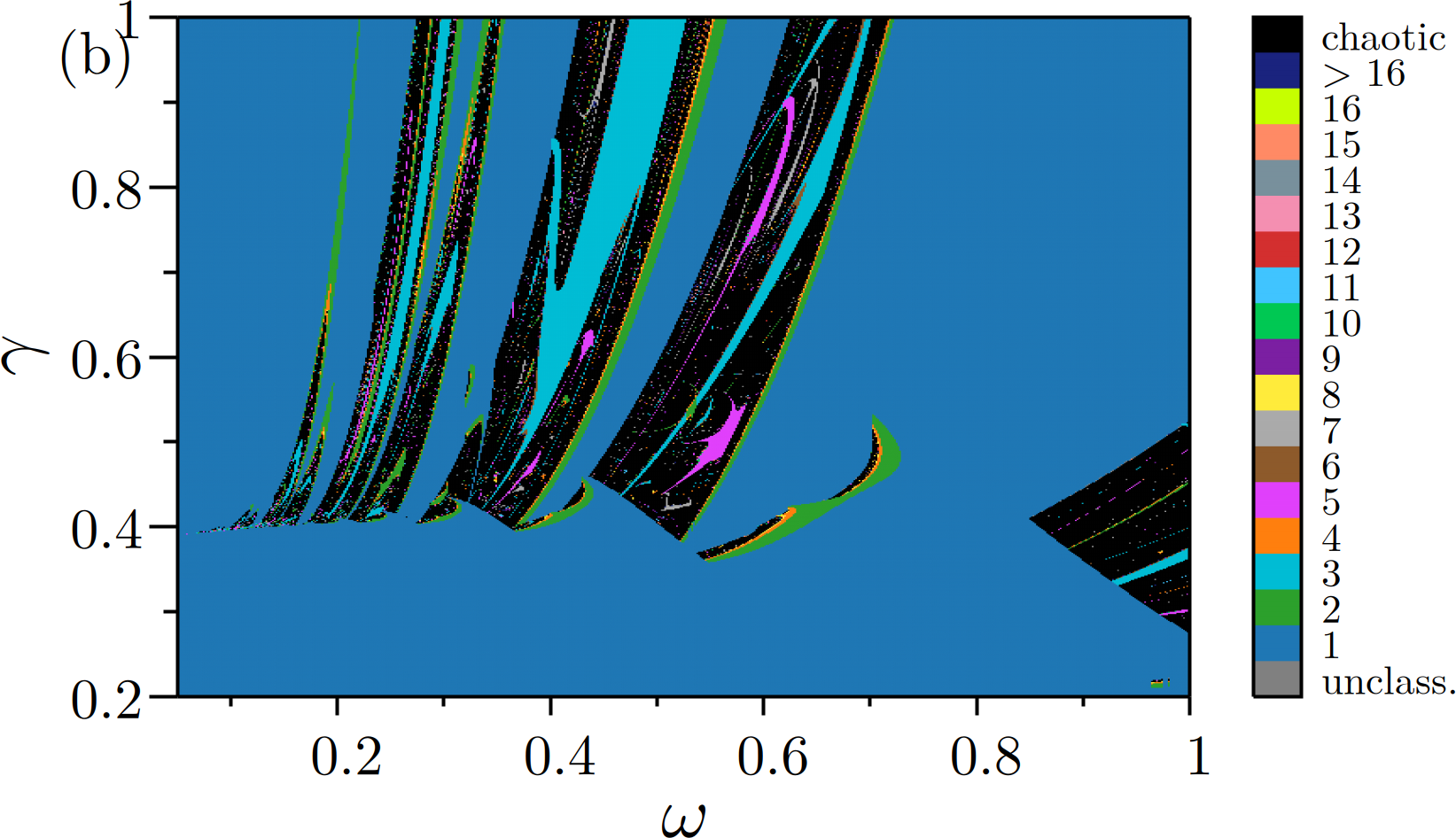}
\includegraphics[width=0.45\textwidth]{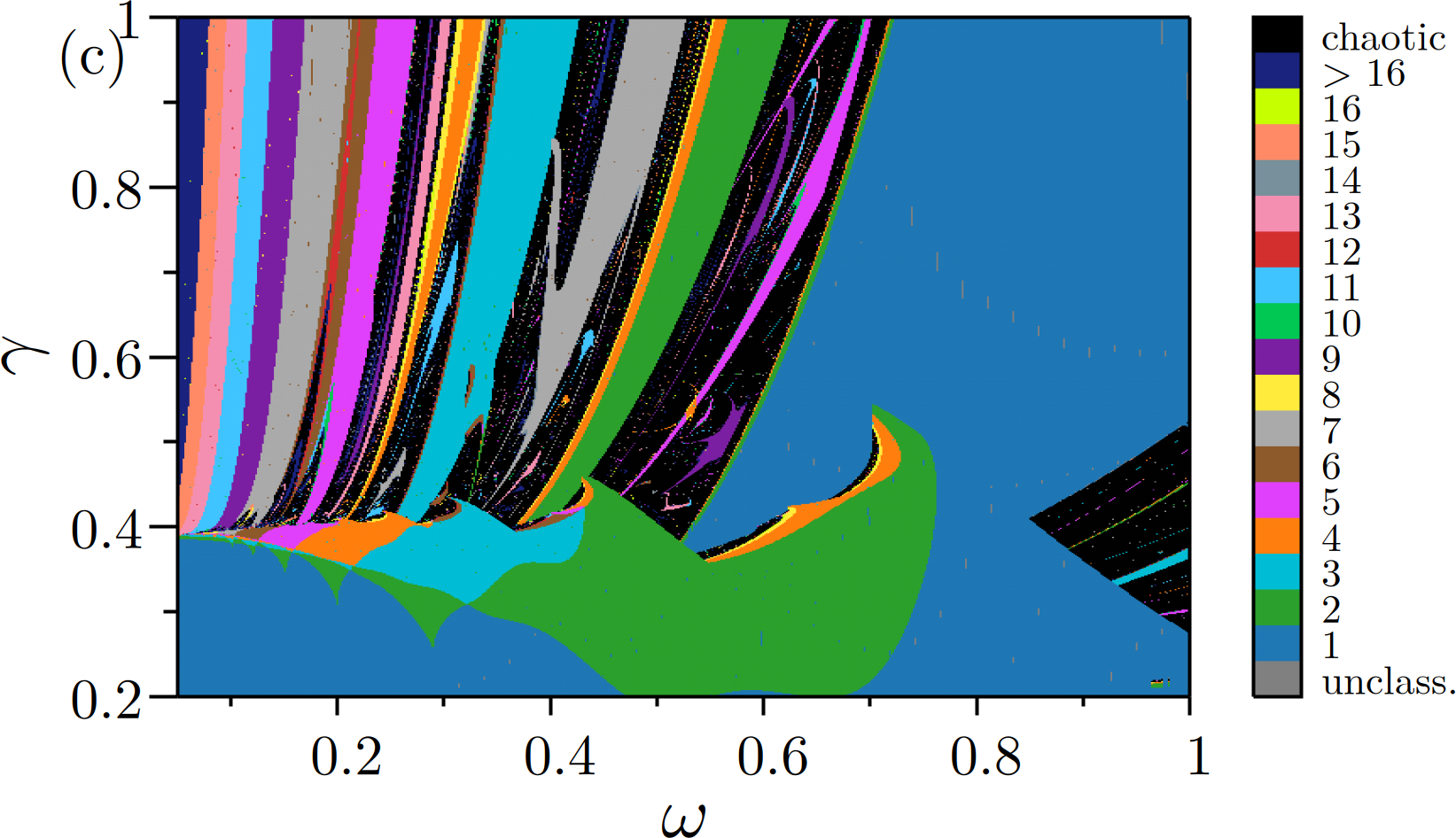}
\caption{\label{fig:origin_maps}
Three complementary dynamical characterizations in the forcing parameter plane
$(\omega, \gamma)$, analogous to those of Fig.~\ref{fig:shrimps_maps},
now for the other region marked with a dashed box in
Fig.~\ref{fig:global}: the low-frequency region close to the origin of
the parameter plane. Panels (a)--(c) show, with the same conventions
and color scales as in Fig.~\ref{fig:shrimps_maps}, the maximum
Lyapunov exponent $\lambda_{\max}$, the orbit order
$k$, and the number of peaks per period,
respectively. Dark-gray dots in panels (b) and (c) mark unclassified
points, as in Fig.~\ref{fig:shrimps_maps}. All panels: $600\times600$ grid over
$\omega\in[0.05,1.0]$ and $\gamma\in[0.2,1.0]$.}
\end{figure}

The scans also allow a global survey of the boundaries of the antiperiodic
windows, which anticipates the two termination scenarios analyzed in
Secs.~\ref{sec:transition} and~\ref{sec:fold}.
In the coarse scan of Fig.~\ref{fig:global} we identified, along
$\gamma$ at fixed $\omega$, all $9\,534$ pairs of adjacent grid points
formed by an antiperiodic orbit of order $k$ and a non-antiperiodic
neighbor. In $65\%$ of them the neighbor is chaotic; in $30\%$ it is a
symmetry-broken periodic orbit of the same order $k$; in
$3\%$ it is a periodic orbit of unrelated order, reached through a jump
between coexisting attractors; and $1\%$ have an unclassified neighbor.
These figures describe only the dynamical class of the
neighboring grid points: adjacency does not by itself identify the
bifurcation responsible for a boundary, particularly in the presence of
multistability and at finite resolution. The pitchfork and the
saddle-node termination demonstrated at high resolution in
Secs.~\ref{sec:transition} and~\ref{sec:fold} are natural candidates for
these two classes of boundaries, but the scan does not establish that
every boundary is produced by the corresponding mechanism.
Only $1.1\%$ ($105$ boundaries) pair an antiperiodic orbit of order $k$
with a periodic orbit of order $2k$, and these involve only $k=1$, $3$
and $5$ (with a single $k=7$ case). At the resolution of this scan,
$\Delta\gamma\approx10^{-2}$, such adjacency does not contradict the
suppression of period doubling: it is compatible with a symmetry-broken
branch of order $k$ that is narrower than the grid step, or whose
residual remains below $\varepsilon_{a}$, and with a
multistability-induced jump between coexisting attractors. Both
alternatives are documented at higher resolution. In the scan of
Fig.~\ref{fig:shrimps_maps}, forty times finer in $\gamma$ ($13\,798$
boundaries), the fraction drops to $0.09\%$ ($13$ boundaries), all of
them in narrow windows of high order, $k\geq7$, while the $k=3$ and $k=5$
boundaries of this type present on the coarse grid in the same region
disappear altogether. In the one-parameter sweep of
Fig.~\ref{fig:bifurcacion_gamma1}, where $\eta$ and $\lambda_{\max}$ are
recorded for every orbit, none of the $51$ boundaries is of this type,
as noted above. In the
low-frequency scan of Fig.~\ref{fig:origin_maps}
($\Delta\gamma\approx1.3\times10^{-3}$, $5\,847$ boundaries) the fraction
is $1.3\%$ ($75$ boundaries); $33$ of them have $k=1$, and most of these
belong to a single band at $\omega\approx0.62$--$0.67$,
$\gamma\approx0.43$, also present in the coarse scan of
Fig.~\ref{fig:global}.

Figure~\ref{fig:shrimps_maps} focuses on the shrimp region. The three
diagnostics offer complementary information. The Lyapunov map [panel~(a)]
delineates the regular--chaotic boundary and reveals the characteristic
shrimp morphology: compact periodic islands, with their two elongated
antennae, embedded in the surrounding chaotic sea. The order map
[panel~(b)] shows that each island corresponds to a periodic window of
definite order $k$, so that the shrimps organize into families of
increasing period as smaller structures accumulate towards the boundary
of the chaotic domain. The peak-count map [panel~(c)] resolves the
internal structure of these windows: within a single shrimp the order $k$
is constant while the number of peaks per period changes across internal
bifurcation lines, revealing the deformation of the waveform that
precedes the loss of stability. The mutual consistency of the three maps
--- every periodic structure visible in $\lambda_{\max}$ reappears with a
definite $k$ and a definite peak count --- confirms the robustness of the
classification, while the information conveyed by each panel remains
complementary: no single diagnostic resolves, by itself, the
regular--chaotic boundary, the order of the orbits, and the internal
structure of their waveforms.

\begin{figure}[htbp]
\centering
\includegraphics[width=\columnwidth]{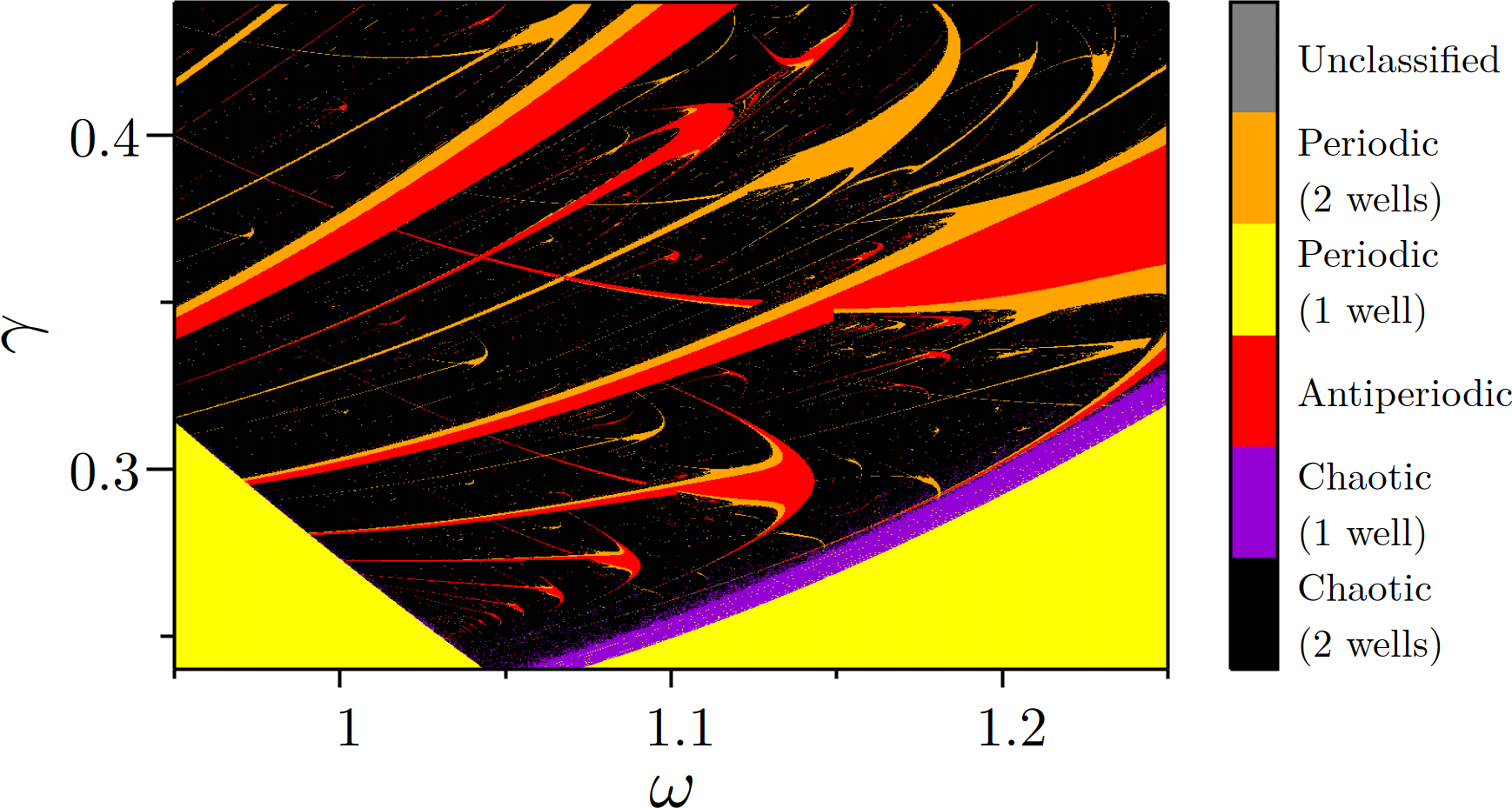}
    \caption{Dynamical regime map
    in the $(\omega, \gamma)$ parameter plane, for the
    region containing the shrimp-shaped periodic structures, marked
    with a dashed box in Fig.~\ref{fig:global}.
    Same grid and parameter ranges as in Fig.~\ref{fig:shrimps_maps}.
    }
    \label{fig:shrimp_regime}
\end{figure}

Figure~\ref{fig:shrimp_regime} supplements these diagnostics with a
fourth map of the same region, which classifies each point jointly by
dynamical regime and by the wells visited. A clear overall trend
emerges: for the smaller forcing amplitudes the system settles
predominantly onto periodic orbits confined to a single well, while
single-well chaos appears as an intermediate situation, in which the
periodic orbits have lost their stability but the trajectory is still
unable to jump from one well to the other. Conversely, for the larger
forcing amplitudes the orbits explore both wells, and these two-well
orbits are predominantly antiperiodic. The antiperiodic orbits, in
turn, explore both wells in every case, consistent with the
well-exchanging symmetry $S$ defined in the Introduction and analyzed in
Sec.~\ref{sec:existence}. This fourth map thus completes the picture
of the dynamics of the system assembled from the three diagnostics of
Fig.~\ref{fig:shrimps_maps}.

Figure~\ref{fig:origin_maps} presents the same three maps for the region
near the origin of the parameter plane. In this low-frequency regime the
plane develops a dense alternation of narrow regular and chaotic tongues,
so that small variations of either parameter suffice to switch the
dynamics between qualitatively different regimes. The order map shows
periodic windows of high $k$, reflecting orbits that close only after
many forcing cycles, while the peak-count map reveals a rapid growth of
the number of peaks per period as $\gamma$ increases at fixed $\omega$
--- the fingerprint of a peak-adding cascade. Note
also that over large portions of panel~(b) the order remains locked at
$k=1$ while the corresponding regions of panel~(c) display a high
number of peaks: the orbit still closes after a single forcing cycle,
but its waveform develops an increasing number of maxima, an
observation in line with the peak-adding phenomenology reported by
Shaw \emph{et al.}~\cite{shaw2015antiperiodic}. As in the
shrimp region, the boundaries traced by the three diagnostics coincide,
and the fine interleaving of tongues anticipates the strong sensitivity
to initial conditions and the multistability-induced discontinuities
documented in the one-parameter diagrams of
Figs.~\ref{fig:bifurcacion_omega13} and~\ref{fig:bifurcacion_gamma1}.

\begin{figure}[htbp]
\centering
\includegraphics[width=\columnwidth]{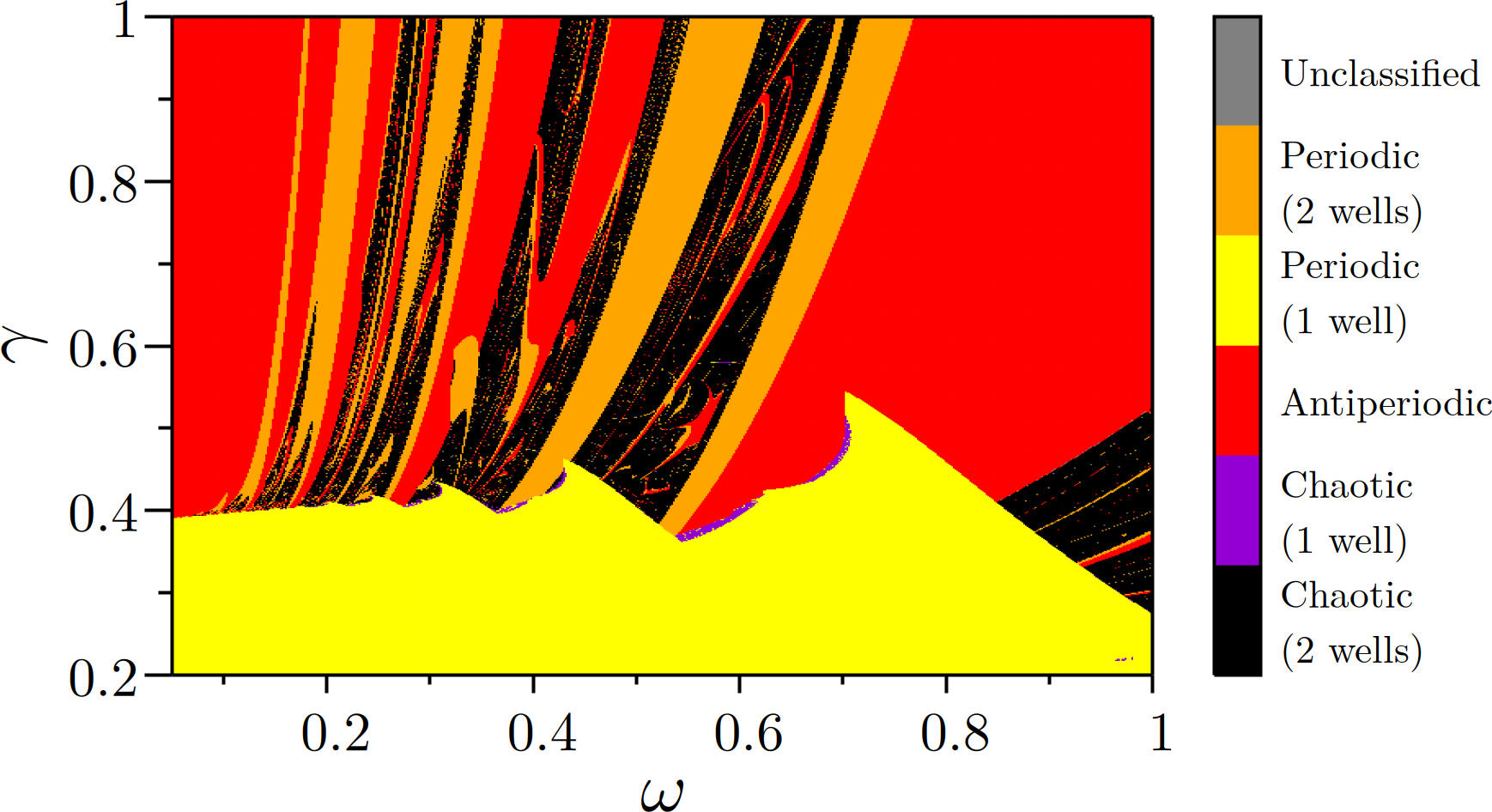}
    \caption{Dynamical regime map, as in
    Fig.~\ref{fig:shrimp_regime}, now for the second region marked with
    a dashed box in Fig.~\ref{fig:global}: the low-frequency region
    close to the origin of the parameter plane. Grid and parameter
    ranges are those of Fig.~\ref{fig:origin_maps}; classification and
    color code are those of Fig.~\ref{fig:shrimp_regime}.}
    \label{fig:origen_regime}
\end{figure}

Figure~\ref{fig:origen_regime} closes the analysis with the
corresponding regime map of the low-frequency region. The general
trends identified in Fig.~\ref{fig:shrimp_regime} persist: orbits
confined to a single well concentrate at the weaker forcing
amplitudes, orbits exploring both wells prevail as $\gamma$ increases,
and the antiperiodic orbits --- again spanning both wells in every
case --- occupy extended regions interleaved with the chaotic tongues.
Altogether, the four analyses applied to each region --- the Lyapunov
exponent, the orbit order, the peak count, and the regime--well
classification --- provide a complete and mutually consistent
characterization of the dynamics across the parameter plane.

\section[Antiperiodicity: origin, existence, conjugate
pairs, and parity selection]{Antiperiodicity: origin, existence, conjugate pairs, and parity
selection}
\label{sec:existence}

For an autonomous system with a single degree of freedom,
$\ddot{x}+V'(x)=0$, the situation is elementary. A nonconstant antiperiodic solution
requires the force to be odd on the range of the solution. Conversely, if the
potential is even, $V(-x)=V(x)$, a periodic orbit is antiperiodic if and only if it
encircles the origin of the phase plane, all other periodic orbits occurring in pairs
exchanged by $x\mapsto-x$; for a double-well potential, this separates the
antiperiodic orbits above the barrier from the pairs of single-well orbits below it.

For $n\geq 2$ degrees of freedom, $\ddot{\mathbf{x}}=-\nabla V(\mathbf{x})$
with $V(-\mathbf{x})=V(\mathbf{x})$, inversion symmetry still guarantees antiperiodic
solutions under standard hypotheses, independently of integrability. If the
linearization at the origin has a pair of eigenvalues $\pm i\omega$ that is
nonresonant with the rest of the spectrum, Lyapunov's center theorem yields a
locally unique one-parameter family of periodic orbits emanating from the
origin~\cite{lamb1998time,montaldi1990existence}; since $\mathbf{x}\mapsto-\mathbf{x}$
preserves this family and the energy, it leaves each orbit invariant, acting on it
as a half-period time shift, so that all these orbits are antiperiodic. For $n=2$,
an additional reflection symmetry of $V$ provides a flow-invariant fixed-point
subspace on which the dynamics reduces to the one-degree-of-freedom case, and hence
antiperiodic orbits over whole ranges of energy, even when the full system is
chaotic. Symmetric periodic orbits of this kind are the subject of an extensive
literature on equivariant and reversible
systems~\cite{lamb1998time,montaldi1990existence}. What does not extend to
$n\geq 2$ is a global criterion identifying the antiperiodic orbits, and bounded
orbits are in general no longer periodic. The inheritance of a discrete symmetry by
chaotic dynamics has been addressed within equivariant dynamics, where a chaotic
attractor may be invariant under a symmetry only as a
set~\cite{chossat1988symmetry,melbourne1993structure,barany1993detecting}. The
periodically driven Duffing oscillator, with one degree of freedom and a
time-periodic forcing, provides a natural setting beyond the autonomous case, whose
rich repertoire of behaviors was illustrated in Sec.~\ref{sec:dynamics}.

The Duffing--Holmes oscillator satisfies the oddness requirement on the
restoring force by
construction: the potential of Eq.~\eqref{eq:potential} is even for
any $\alpha$ and $\beta$. The evenness of $V$, and hence the invariance under $S$, holds
therefore also in the single-well case
$\alpha>0$~\cite{sato1983universal,novak1982transition}; the double-well
geometry, $\alpha<0<\beta$, is a choice of regime rather than part of the
symmetry requirement, and the interpretation of $S$ as an exchange of the two
wells, used throughout this work, is specific to it. Once
the periodic forcing is introduced, however, the system is no longer
autonomous and the symmetry of the potential alone no longer suffices;
the drive itself must be compatible with the sought symmetry. For
Eq.~\eqref{eq:duffing} driven by $\gamma\cos(\omega t)$ this is the case:
the forcing is antiperiodic with antiperiod $\pi/\omega$, so that
solutions sharing this antiperiodic structure are \textit{a priori}
compatible with the symmetry of the equations of motion. Compatibility is
a necessary condition --- in systems whose forcing breaks the odd
symmetry, antiperiodic steady states are generically destroyed --- but it
does not guarantee existence.

The degree of nonlinearity, controlled by $\beta$, also plays a decisive
role. For $\beta = 0$ the system is linear and the only periodic (and
antiperiodic) solution is the one locked at the forcing frequency
(which, for $\alpha<0$, is moreover unstable).
Nonzero $\beta$ opens the possibility of subharmonic and ultraharmonic
resonances, and it is within this broader landscape that the hierarchy of
antiperiodic solutions with multiple peaks per antiperiod can emerge.

Taken together, these considerations show that Eq.~\eqref{eq:duffing}
satisfies the necessary conditions for antiperiodic solutions to exist,
but do not by themselves establish where in parameter space such
solutions occur, nor how they relate to the periodic and chaotic regions
identified in Sec.~\ref{sec:dynamics}. The distribution of the
dynamical regimes over the parameter plane was presented in the regime
maps of Figs.~\ref{fig:shrimp_regime} and~\ref{fig:origen_regime},
which located the antiperiodic regions and the wells visited by each
orbit. The remainder of this section accounts for the two features
highlighted there: antiperiodic orbits occupy well-defined regions
rather than isolated points, and every antiperiodic orbit explores
both potential wells, as expected for a symmetry that exchanges the
wells.

\begin{figure}[htbp]
\centering
\includegraphics[width=\columnwidth]{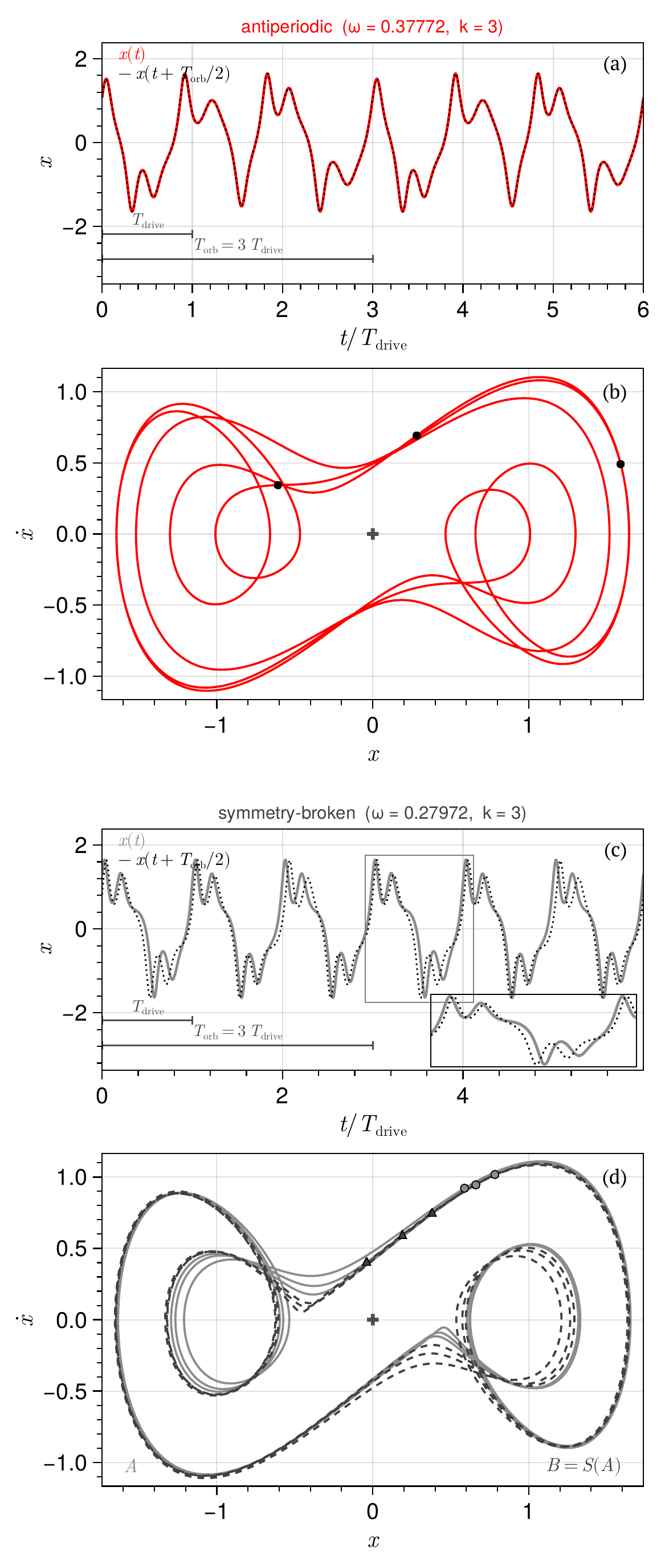}
\caption{Antiperiodic versus symmetry-broken orbits at $\gamma=0.5$.
(a),(b)~Time series and phase portrait of the antiperiodic orbit at
$\omega=0.37772$: $x(t)$ (red) coincides with
$-x(t+T_{\mathrm{orb}}/2)$ (dotted black).
(c),(d)~Same for the symmetry-broken orbit at $\omega=0.27972$: the two
curves no longer coincide [inset: region of maximal separation]; its
conjugate twin $B=S(A)$ is shown dashed in~(d).
Brackets mark $T_d$ and
$T_{\mathrm{orb}}=3\,T_d$; dots and triangles mark the
stroboscopic sections; the cross marks the origin.}
    \label{fig:mecanismo}
\end{figure}

The dynamical origin of these regions is the half-period shift symmetry
$S$ of the equations of motion, defined in the Introduction, under
which Eq.~\eqref{eq:duffing} is invariant. A periodic orbit can respond
to this symmetry in two ways, illustrated in Fig.~\ref{fig:mecanismo}
with two orbits at $\gamma=0.5$ that lock to the drive with the same odd
ratio, $T_{\mathrm{orb}}=3\,T_d$. The orbit can be
individually invariant under $S$, in which case it is antiperiodic: for
the orbit at $\omega=0.37772$, the time series reproduces its own
negative after half an orbital period,
$x(t)=-x(t+T_{\mathrm{orb}}/2)$ [Fig.~\ref{fig:mecanismo}(a)], and the
phase portrait is point-symmetric through the origin
[Fig.~\ref{fig:mecanismo}(b)]. Alternatively, the orbit can break the
symmetry: for the orbit at $\omega=0.27972$, $x(t)$ and
$-x(t+T_{\mathrm{orb}}/2)$ no longer coincide
[Fig.~\ref{fig:mecanismo}(c)], so the orbit is periodic but not
antiperiodic. In this case the symmetry is restored globally by the
coexistence of a conjugate twin $B=S(A)$, the point reflection of the
original orbit through the origin [Fig.~\ref{fig:mecanismo}(d)]: the
pair jointly recovers the invariance lost by each orbit individually.
Antiperiodicity is thus not an accidental property of particular
waveforms but the orbit-level manifestation of a symmetry of the
system, and its absence in a given periodic window signals spontaneous
symmetry breaking.

This dichotomy has a precise, model-independent formulation in
the theory of symmetric periodically forced systems developed by Swift and
Wiesenfeld~\cite{swift1984suppression}. In first-order form [cf.\
Eq.~\eqref{eq:system}], the symmetry $S$ is equivalent to
$\mathbf{f}(\mathbf{u},t)=-\mathbf{f}(-\mathbf{u},t+T_d/2)$, under which the
drive-period Poincar\'e map factors as $P=\tilde{P}^{2}$, with $\tilde{P}$ the
half-period map composed with the inversion
$\mathbf{u}\mapsto-\mathbf{u}$~\cite{swift1984suppression,lamb1998time}. The
cycle structure of $\tilde{P}$ then encodes the dichotomy for orbits of
arbitrary order: an $m$-cycle of $\tilde{P}$ is a single antiperiodic orbit of
order $k=m$ if $m$ is odd, and a conjugate pair of orbits of order $k=m/2$,
exchanged by $\tilde{P}$, if $m$ is even~\cite{melbourne1993structure}. The
dichotomy extends to arbitrary attractors: every attractor of $P$ is either
$\tilde{P}$-invariant or one member of a conjugate pair with
$A\cap\tilde{P}(A)=\emptyset$~\cite{chossat1988symmetry,melbourne1993structure}.
Because $S$ has no fixed points, invariance amounts to antiperiodicity for a
periodic attractor, whereas for a chaotic attractor it constrains only the
attractor as a set and its natural
measure~\cite{chossat1988symmetry,barany1993detecting}; hence confinement to a
single well implies broken symmetry, but visiting both wells does not imply
$S$-invariance. The factorization also accounts for the suppression of period
doubling~\cite{swift1984suppression}: a Floquet multiplier $-1$ of
$P^{k}=(\tilde{P}^{k})^{2}$ would require the real matrix $D\tilde{P}^{k}$ to
have the eigenvalue pair $\pm i$, and hence $-1$ as a double multiplier of
$P^{k}$, which is incompatible with $\det DP^{k}=e^{-\delta kT_d}<1$ [cf.\
Eq.~\eqref{eq:trace}]. The parity selection rule confirmed numerically below is
thus a consequence of the factorization alone, independent of the particular
nonlinearity or potential shape.

In the classification of Lamb and
Roberts~\cite{lamb1998time}, $S$ is a space--time symmetry (an equivariance)
rather than a reversing symmetry; indeed, for $\delta>0$ no reversing symmetry
can exist, since it would have to map the uniform contraction of phase-space
volumes [cf.\ Eq.~\eqref{eq:trace}] onto an expansion. The distinction is
essential: limit sets invariant under a reversing symmetry cannot be
asymptotically stable, whereas $S$-invariant attractors,
such as the antiperiodic orbits studied here, are not excluded
\cite{lamb1998time,lamb1998ksymmetry}.

\begin{figure}[htbp]
\centering
\includegraphics[width=\columnwidth]{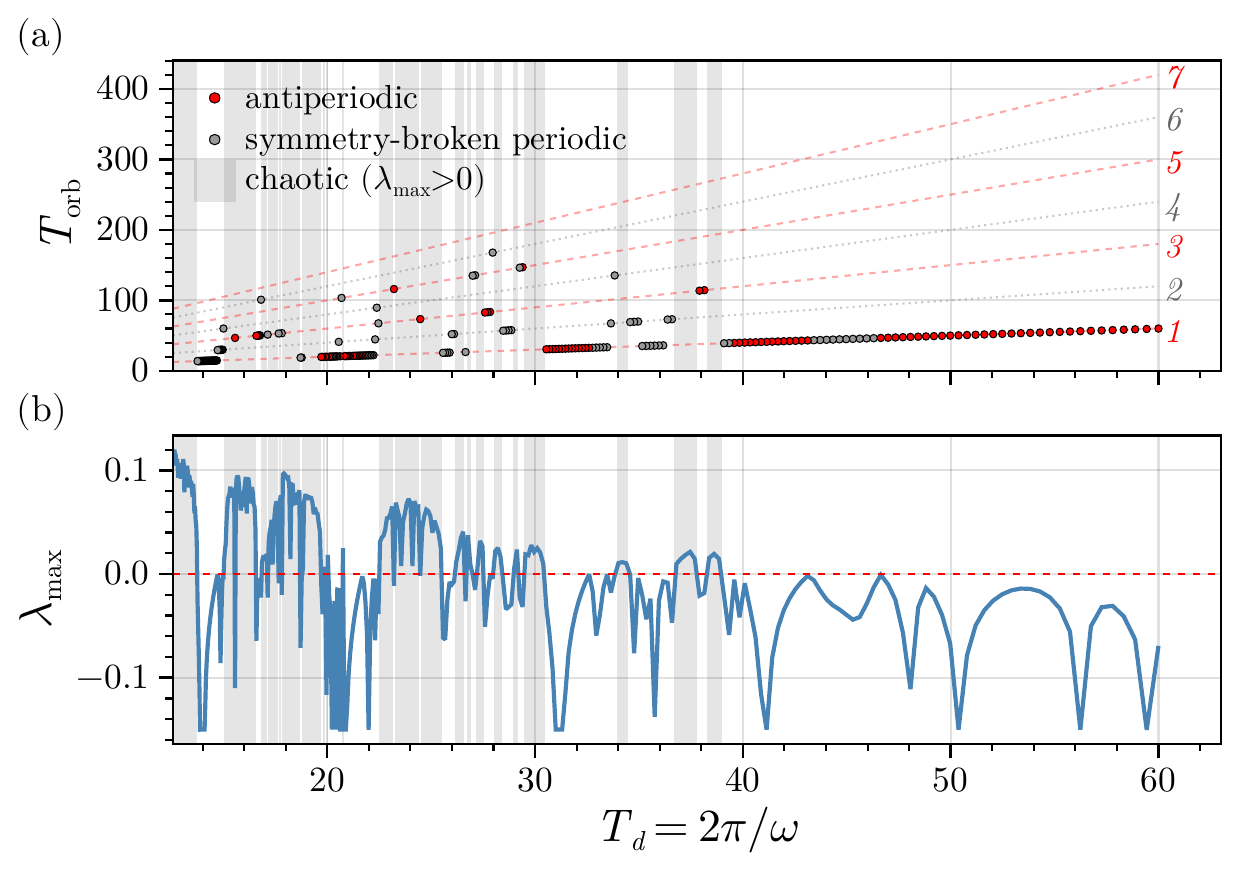}
 \caption{\label{fig:sombreada}
Parity selection rule at $\gamma=0.5$.
The frequency is swept from $\omega=2\pi/60$ to $\omega=0.5$ with
step $\Delta\omega=10^{-3}$.
(a)~Orbital period $T_{\mathrm{orb}}$ versus drive period
$T_d=2\pi/\omega$. Reference lines
$T_{\mathrm{orb}}=k\,T_d$: $k$ odd (dashed red), $k$ even (dotted
gray). red circles: antiperiodic orbits; gray circles: symmetry-broken
periodic orbits. Shaded vertical bands: chaotic windows
($\lambda_{\max}>0$).
(b)~Maximum Lyapunov exponent over the same sweep.
marks $\lambda_{\max}=0$.
}
\end{figure}

Invariance under $S$ imposes, in addition, a selection rule on the
locking ratio $k$. An $S$-invariant orbit of period $kT_d$ satisfies
$x(t+kT_d/2)=-x(t)$. This half-\emph{orbit}-period relation must be
distinguished from the half-\emph{drive}-period shift entering $S$: the two
are compatible because, for $T_{\mathrm{orb}}=(2n+1)T_d$,
$T_{\mathrm{orb}}/2=nT_d+T_d/2$, and the drive is $T_d$-periodic. If $k$ were even, $kT_d/2$ would be an integer
number of drive periods, so that $-x(t)$ would solve
Eq.~\eqref{eq:duffing} with the \emph{same} forcing as $x(t)$; adding
the two equations, whose left-hand sides are odd in $x$, gives
$\gamma\cos(\omega t)\equiv0$, a contradiction for $\gamma\neq0$. Hence
antiperiodicity requires $k$ odd, in agreement with the cycle structure
of $\tilde{P}$ derived above. Figure~\ref{fig:sombreada} confirms this parity rule
along a sweep at $\gamma=0.5$, in which the attractor is followed by
continuation in $\omega$ from the common reference state
$\mathbf{u}_{0}$ (Appendix~\ref{subsec:scan}) at
$\omega=2\pi/60$ up to $\omega=0.5$, with step $\Delta\omega=10^{-3}$, and the period of
each asymptotic orbit is obtained from the first return of the
stroboscopic map. Every detected orbit locks to an integer multiple of
the drive, $T_{\mathrm{orb}}=k\,T_d$, so that in the
$(T_d,T_{\mathrm{orb}})$ plane the points accumulate on
the straight lines $T_{\mathrm{orb}}=k\,T_d$. Antiperiodic orbits fall
exclusively on the lines of odd $k$, as required by the selection rule,
whereas symmetry-broken periodic orbits populate both parities. The
locked windows are separated by chaotic intervals, shaded in both
panels, where $\lambda_{\max}>0$ and no period is defined; the Lyapunov
exponent of panel~(b) confirms that every unlocked interval is genuinely chaotic
rather than an unresolved periodic window. The parity rule is a
necessary condition only: odd-$k$ windows may host either antiperiodic
or symmetry-broken orbits, and it is the competition between these two
possibilities that organizes the fine structure of the regime map of
Fig.~\ref{fig:origen_regime}.

\section[Transition from antiperiodicity to broken symmetry: the
pitchfork]{Transition from antiperiodicity to broken
symmetry: the pitchfork}
\label{sec:transition}

\begin{figure}[htbp]
\centering
\includegraphics[width=\columnwidth]{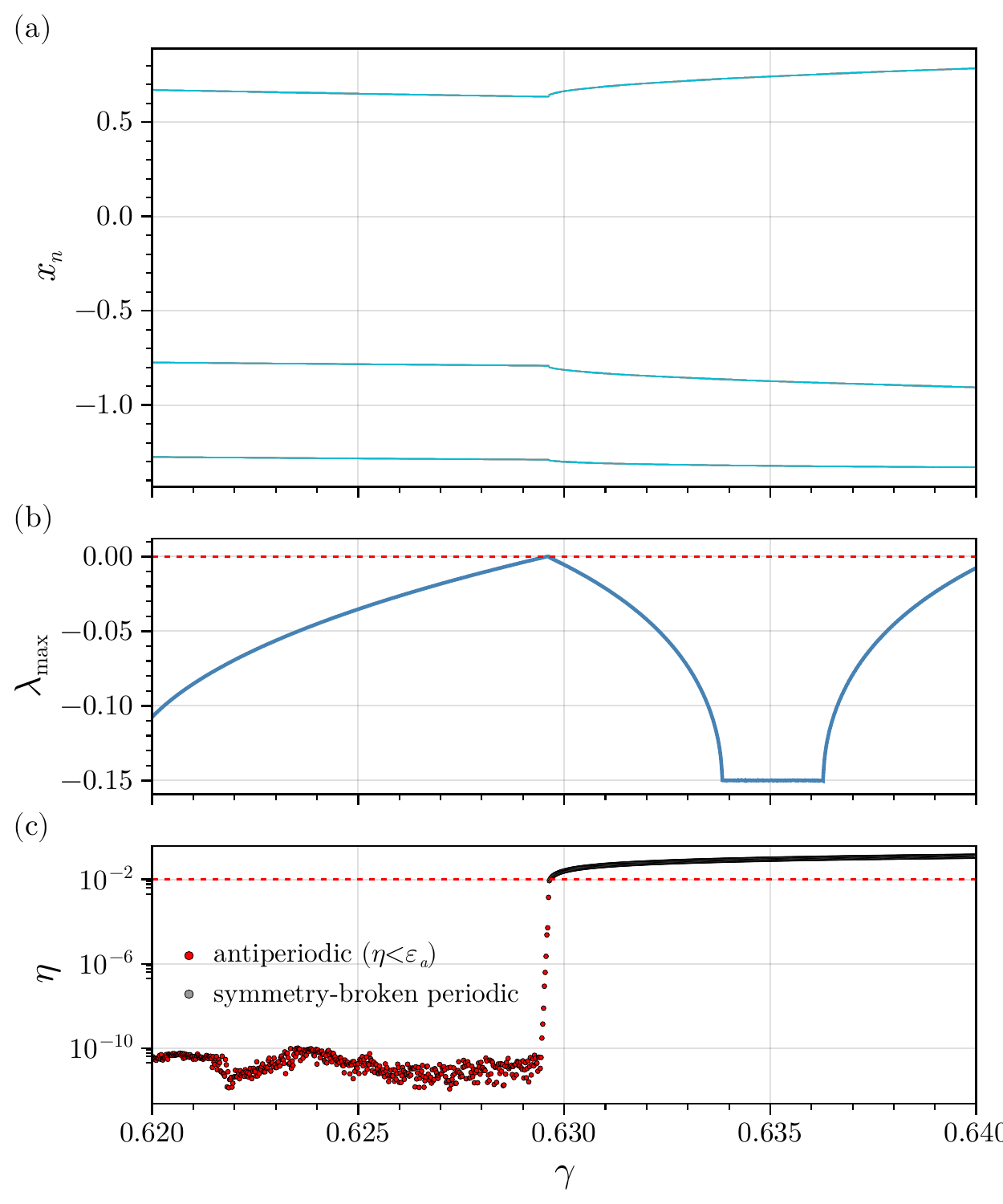}
\caption{\label{fig:zoom_omega13}%
Magnification of the bifurcation diagram of
Fig.~\ref{fig:bifurcacion_omega13} across the upper exit of the
antiperiodic window at $\omega=1.3$: $\gamma\in[0.62,0.64]$, scanned
with step $\Delta\gamma=2\times10^{-5}$.
(a)~Stroboscopic section $x_n$: the orbit remains periodic with
constant order $k=3$ (cyan) over the whole interval, with neither
period doubling nor chaotic windows.
(b)~Maximum Lyapunov exponent $\lambda_{\max}$, which stays negative
throughout and approaches zero at $\gamma_{c}$
(Fig.~\ref{fig:pitchfork}).
(c)~Antiperiodicity residual $\eta$ [symbols, threshold, and
shading as in Fig.~\ref{fig:bifurcacion_omega13}(c)]. Across the
transition $\eta$ climbs by about nine orders of magnitude, from the
numerical floor to values above the threshold.}
\end{figure}

Having established that every periodic window must host
either an antiperiodic orbit or a symmetry-broken conjugate pair, we now
examine how the system passes from one situation to the other, i.e.\
what happens in the immediate vicinity of the transition from
antiperiodicity to broken symmetry. The bifurcation diagram of
Fig.~\ref{fig:bifurcacion_omega13} provides a natural case study: at
$\omega=1.3$ the antiperiodic window ends near $\gamma\approx0.63$,
where the residual of panel~(c) leaves the numerical floor and crosses
the threshold $\varepsilon_{a}$ while the orbit remains periodic.
Figure~\ref{fig:zoom_omega13} magnifies this exit over the interval
$\gamma\in[0.62,0.64]$, resolved with a step
$\Delta\gamma=2\times10^{-5}$, twenty-five times finer than that of the
full sweep. The magnification shows that the transition involves
neither chaos nor a change of order: the stroboscopic section retains a
constant order $k=3$ (cyan) across the whole interval, and
$\lambda_{\max}$ remains negative, approaching zero precisely where
$\eta$ climbs by some nine orders of magnitude. The
antiperiodic orbit thus does not lose stability to chaos or period
doubling; it sheds its symmetry, in a bifurcation analyzed
quantitatively next.

\begin{figure}[htbp]
\centering
\includegraphics[width=\columnwidth]{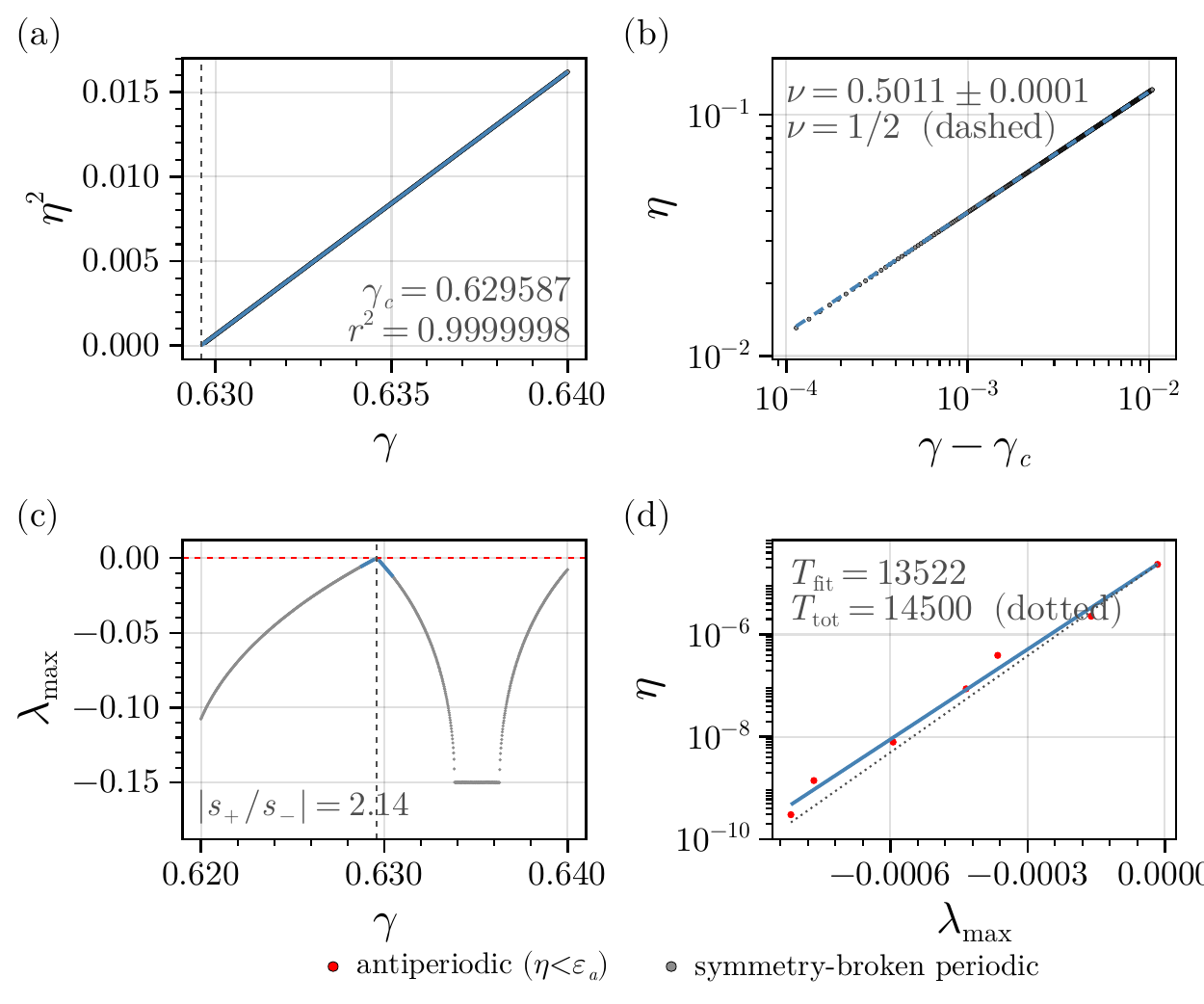}
\caption{\label{fig:pitchfork}%
Supercritical symmetry-breaking bifurcation of the antiperiodic orbit at
$\omega=1.3$, from the sweep of Fig.~\ref{fig:zoom_omega13} (order $k=3$
throughout the interval shown, with no period doubling and no chaotic
window).
(a)~Squared antiperiodicity residual $\eta^{2}$
[Eq.~\eqref{eq:antip_residual}] on the symmetry-broken branch: the
linear growth (solid line) locates the critical amplitude
$\gamma_{c}=0.629587$ (vertical dashed line).
(b)~Same data on logarithmic scales:
$\eta=1.247\,(\gamma-\gamma_{c})^{\nu}$ with fitted
 $\nu=0.501\pm0.002$ (the
uncertainty displayed in the panel is the formal fitting error; see
text); the dashed line has the normal-form slope
$\nu=1/2$. Symbols as in Fig.~\ref{fig:bifurcacion_omega13}(c).
(c)~Maximum Lyapunov exponent: $\lambda_{\max}$ vanishes at the same
$\gamma_{c}$, and the side slopes $s_{\pm}=d\lambda_{\max}/d\gamma$
have a ratio close to the value $2$ required by the pitchfork normal
form (see text) (red: antiperiodic branch; gray: symmetry-broken
branch). The plateau at $\lambda_{\max}=-\delta/2$
signals the interval where the two Floquet multipliers form a
complex-conjugate pair, so that $\lambda_{1}=\lambda_{2}=-\delta/2$
follows from the constant contraction rate of Eq.~\eqref{eq:trace}.
(d)~Sub-critical residual against $\lambda_{\max}$ over the last points
of the $S$-invariant branch: the exponential dependence
$\eta\propto\exp(\lambda_{\max}T_{\mathrm{tot}})$, with fitted slope
$T_{\mathrm{fit}}$ (solid line) compared with the slope fixed at the
total integration time per parameter value, $T_{\mathrm{tot}}$ (dotted
line), identifies the apparent pre-critical growth
of $\eta$ as an unrelaxed transient, critical slowing down, rather
than a physical asymmetry.}
\end{figure}

The residual $\eta$ of Eq.~\eqref{eq:antip_residual} is, up to a factor
of two, the amplitude of the $S$-odd component of the orbit normalized
by the orbit amplitude, and therefore serves as an order parameter for
the breaking of the symmetry $S$. Figure~\ref{fig:pitchfork} deepens
the analysis of the transition magnified in Fig.~\ref{fig:zoom_omega13},
recasting the same data in terms of this order parameter. With chaos
and period doubling already excluded there, the signatures of the
transition are those of a clean supercritical pitchfork.
The squared residual grows linearly with the distance to the critical
point, $\eta^{2}\propto\gamma-\gamma_{c}$ with $r^{2}>0.999999$ [panel
(a)], yielding $\gamma_{c}=0.629587$ and the amplitude law
$\eta=1.247\,(\gamma-\gamma_{c})^{1/2}$; a direct fit of the exponent
gives $\nu=0.501\pm0.002$
[panel (b)], the uncertainty being
dominated by the choice of fitting window rather than by statistical
error (the formal fitting error, $\pm0.0001$, is the
value displayed in the panel).

The maximum Lyapunov exponent provides two independent checks
[panel (c)]: it vanishes at the same $\gamma_{c}$, as expected when a
real Floquet multiplier crosses the unit circle, and the slopes
$s_{\pm}=d\lambda_{\max}/d\gamma$ measured on the two sides of the
transition satisfy $|s_{+}/s_{-}|=2.14$ for linear fits over a half-width $|\gamma-\gamma_{c}|<8\times10^{-4}$,
to be compared with the value $2$ imposed by the
pitchfork normal form, in which the critical multiplier behaves as
$1+a(\gamma-\gamma_{c})$ on the symmetric branch and as
$1-2a(\gamma-\gamma_{c})$ on each asymmetric branch. Since the slope
ratio alone does not discriminate a pitchfork from a period doubling
--- the period-doubling normal form yields the same factor of two ---
the identification rests on the conserved order $k$ and on the growth
of the $S$-odd amplitude itself: the antiperiodic orbit bifurcates into
a conjugate pair $A$, $B=S(A)$ of the same order, realizing at the
level of a single orbit the dichotomy of Fig.~\ref{fig:mecanismo}. This
constitutes a direct, quantitative instance of the general result of
Swift and Wiesenfeld~\cite{swift1984suppression}, who discuss precisely a
symmetric period-$3T$ orbit within a periodic window that breaks its symmetry
before the subsequent period-doubling cascade: the antiperiodic
orbit does not period-double; it first sheds its symmetry through a
supercritical pitchfork with the mean-field exponent $\nu=1/2$, and
any subsequent period doubling can only take place on the
symmetry-broken branch.

Panel (d) of Fig.~\ref{fig:pitchfork} documents a numerical subtlety
that accompanies every such transition. Below $\gamma_{c}$ the measured
residual is not identically zero: within
$\Delta\gamma\approx1.3\times10^{-4}$ of the critical point it climbs by
some six orders of magnitude above the round-off floor, and it reaches the
detection threshold $\varepsilon_{a}$ just beyond $\gamma_{c}$. This ramp
is not a physical asymmetry of the attractor but the unrelaxed transient of
the critical mode. Each increment of the control parameter displaces the
$S$-invariant orbit, so that the state inherited from the previous
parameter value carries a small component, of relative size
$\eta_{0}=O(\Delta\gamma)$, along the critical $S$-odd Floquet direction;
over the total integration time spent at each parameter value,
$T_{\mathrm{tot}}=T_{\mathrm{tr}}+T_{\mathrm{int}}$, the sum of the
discarded transient and of the recorded window
(Appendix~\ref{subsec:integration}), this component decays
as $\exp(\lambda_{\max}T_{\mathrm{tot}})$. Writing
$\eta_{\mathrm{meas}}\approx\eta_{0}\exp(\lambda_{\max}T_{\mathrm{tot}})$
predicts a linear relation between $\ln\eta$ and $\lambda_{\max}$ with
slope $T_{\mathrm{tot}}$. The fitted slope, $(1.35\pm0.06)\times10^{4}$,
agrees with the nominal value $1.45\times10^{4}$ to within $7\%$, and the
fitted prefactor, $\eta_{0}\approx3\times10^{-5}$, is indeed comparable to
the step $\Delta\gamma=2\times10^{-5}$. Critical slowing down thus
displaces the apparent onset of symmetry breaking by an amount that
shrinks only as $1/T_{\mathrm{tot}}$; the points within this ramp were
excluded from the fits above, and the effect should be kept in mind
whenever antiperiodicity is diagnosed by thresholding $\eta$ in the
vicinity of a symmetry-breaking point.

\section[Termination of antiperiodicity with the symmetry intact: the
saddle-node bifurcation]{Termination of antiperiodicity
with the symmetry intact: the saddle-node bifurcation}
\label{sec:fold}

\begin{figure}[htbp]
\centering
\includegraphics[width=\columnwidth]{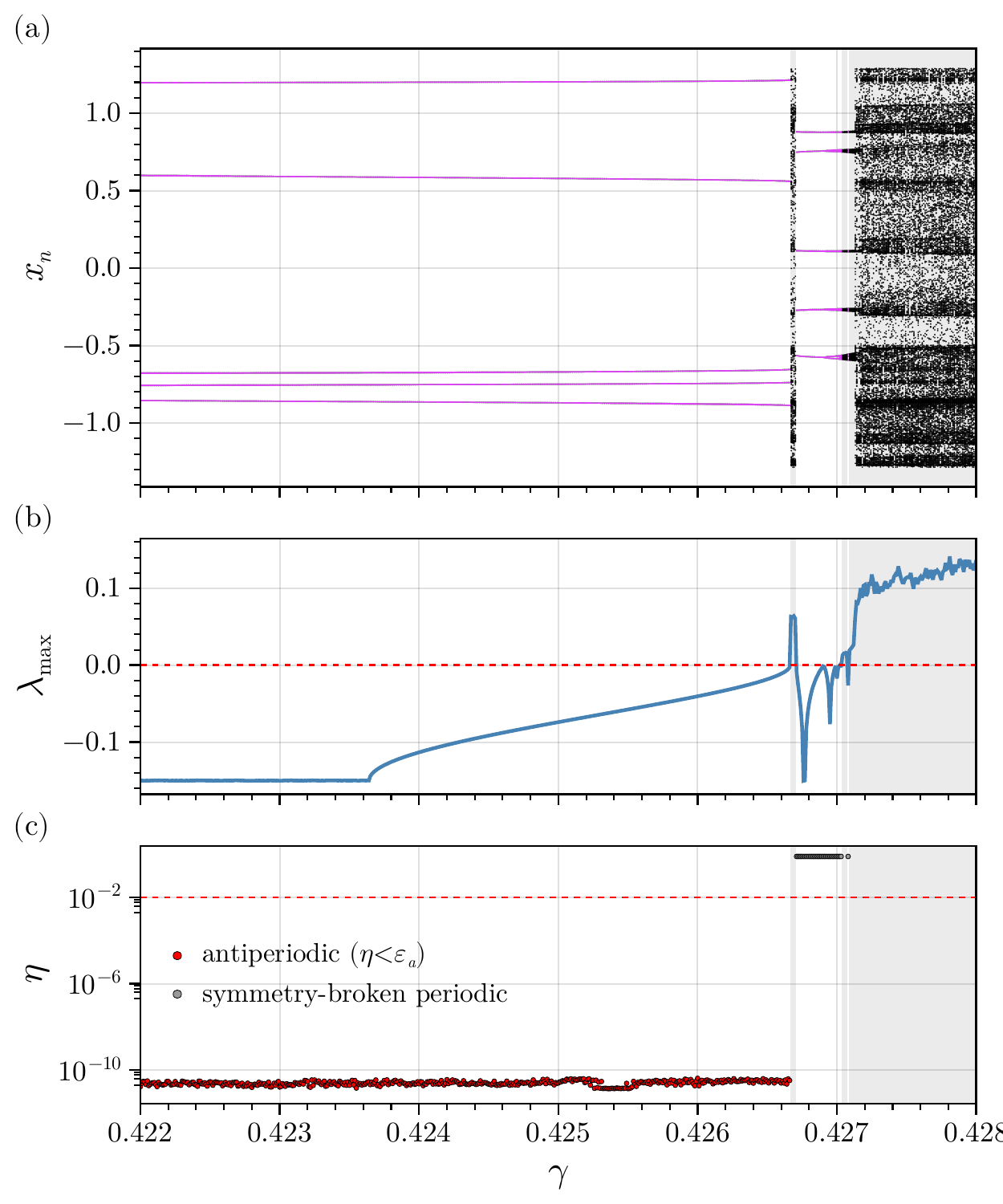}
\caption{\label{fig:zoom_fold}%
Magnification of the bifurcation diagram of
Fig.~\ref{fig:bifurcacion_omega13} across the upper end of the antiperiodic
window of order $k=5$ at $\omega=1.3$: $\gamma\in[0.422,0.428]$, scanned
with step $\Delta\gamma=10^{-5}$.
(a)~Stroboscopic section $x_n$: the five branches of the antiperiodic orbit
end abruptly at $\gamma\approx0.42666$; beyond a narrow chaotic interval
(shaded) the continuation follows one member of a conjugate pair of
asymmetric orbits, also of order $k=5$ but located elsewhere in the section,
and then its period-doubling cascade and the ensuing chaotic attractors.
(b)~Maximum Lyapunov exponent $\lambda_{\max}$: it leaves the plateau
$-\delta/2$ at $\gamma\approx 0.42364$, where the Floquet multipliers of the
orbit become real, and rises to zero at the end of the antiperiodic branch.
(c)~Antiperiodicity residual $\eta$ [symbols, threshold, and shading as in
Fig.~\ref{fig:bifurcacion_omega13}(c)]: it remains at the numerical floor up
to the last antiperiodic point and then changes discontinuously, by more than
ten orders of magnitude, in contrast with the continuous growth of
Fig.~\ref{fig:zoom_omega13}(c).}
\end{figure}

\begin{figure}[htbp]
\centering
\includegraphics[width=\columnwidth]{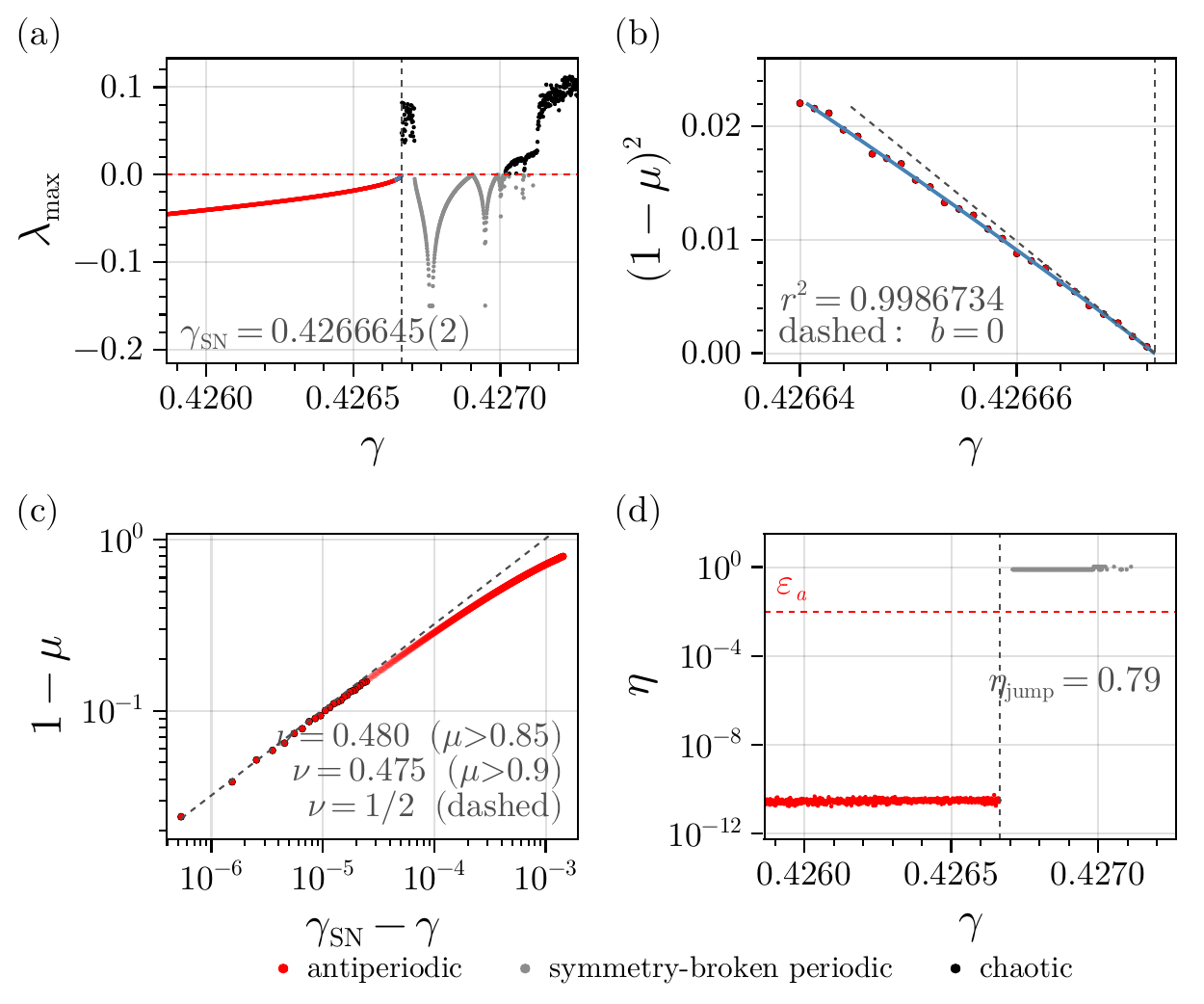}
\caption{\label{fig:fold}%
Saddle-node (fold) termination of the antiperiodic orbit of
order $k=5$ at $\omega=1.3$, from a continuation sweep with step
$\Delta\gamma=10^{-6}$. red: antiperiodic ($\eta<\varepsilon_{a}$); gray:
symmetry-broken periodic; black: chaotic.
(a)~Maximum Lyapunov exponent: on the antiperiodic branch $\lambda_{\max}$
reaches zero with a vertical tangent at
$\gamma_{\mathrm{SN}}=0.4266645(2)$ (vertical dashed line); the solid line is
the fit of panel (b). Beyond the fold the trajectory visits a narrow chaotic
interval, a pair of asymmetric $k=5$ orbits born in a saddle-node bifurcation
of their own, their period-doubling cascade, and the ensuing chaotic
attractors.
(b)~Squared distance of the dominant Floquet multiplier from unity,
$\mu=\exp(\lambda_{\max}kT_d)$, for the $25$ points with $\mu>0.85$: fit of
$\gamma=\gamma_{\mathrm{SN}}-a(1-\mu)^{2}-b(1-\mu)^{3}$ (solid line) and its
leading term alone (dashed line), which is the saddle-node normal form.
(c)~Same data on logarithmic scales, including the points with
$0.2<\mu\le0.85$ excluded from the fit (light symbols); the dashed line is the leading term
of panel (b), of slope $\nu=1/2$, with no adjustable parameter; the
logarithmic slopes fitted over two windows are indicated.
(d)~Antiperiodicity residual $\eta$ [Eq.~\eqref{eq:antip_residual}]: it
remains at the numerical floor up to $\gamma_{\mathrm{SN}}$, with no
pre-critical ramp, and then jumps discontinuously to
$\eta_{\mathrm{jump}}=0.79$ on the asymmetric branch; $\eta$ is undefined in
the intervening chaotic interval. Horizontal dashed line: detection
threshold $\varepsilon_{a}$.}
\end{figure}

The pitchfork is not the only way in which an antiperiodic
window can end. Since a multiplier $-1$ is excluded, and a complex-conjugate
pair of multipliers, whose modulus is fixed at $e^{-\delta kT_d/2}<1$ by
Eq.~\eqref{eq:trace}, can never reach the unit circle, an $S$-invariant orbit of
order $k$ can lose stability only when a real
multiplier of $P^{k}=(\tilde{P}^{k})^{2}$ reaches $+1$, and the factorization
leaves two possibilities, according to the eigenvalue of $D\tilde{P}^{k}$
responsible for it. If that eigenvalue is $-1$, the critical direction is
reversed by $\tilde{P}^{k}$, i.e.\ it is $S$-odd: the orbit persists, becomes
unstable, and sheds a conjugate pair, which is the pitchfork of
Figs.~\ref{fig:zoom_omega13} and~\ref{fig:pitchfork}. If it is $+1$, the
critical direction is $S$-even and the generic outcome is a saddle-node
(fold) bifurcation of $\tilde{P}^{k}$ itself: the antiperiodic orbit collides
with a saddle that is likewise $S$-invariant, and both disappear without any
breaking of the symmetry. These are precisely the two
generic local codimension-one bifurcations available to a symmetric
orbit in the classification of Swift and
Wiesenfeld~\cite{swift1984suppression}: the symmetry-breaking
(pitchfork) bifurcation and the saddle-node bifurcation, period doubling
being excluded. The $k=5$ window of
Fig.~\ref{fig:bifurcacion_omega13} provides an instance of this second
scenario: its upper end is magnified in Fig.~\ref{fig:zoom_fold} and analyzed
quantitatively in Fig.~\ref{fig:fold}.

Figure~\ref{fig:zoom_fold} displays the interval
$\gamma\in[0.422,0.428]$, scanned with step $\Delta\gamma=10^{-5}$, fifty
times finer than that of Fig.~\ref{fig:bifurcacion_omega13}. Up to
$\gamma\approx0.42666$ the stroboscopic section consists of five branches
that vary smoothly with $\gamma$, and the residual stays at the numerical
floor, $\eta\sim10^{-11}$, without any trend. The only diagnostic that
anticipates the end of the window is $\lambda_{\max}$. While the two Floquet
multipliers of the orbit are complex conjugate, their common modulus is fixed
by the dissipation and $\lambda_{\max}$ sits on the plateau $-\delta/2$; at
$\gamma\approx 0.42364$ they collide on the real axis, and from there on
$\lambda_{\max}$ increases monotonically up to zero. What follows differs
qualitatively from Fig.~\ref{fig:zoom_omega13}. There, the symmetry-broken
branches emerge continuously from the antiperiodic one; here, the five
antiperiodic branches end abruptly, and the periodic orbit found beyond a
narrow chaotic interval, again of order $k=5$ but with $\eta$ of order unity,
occupies positions in the section unrelated to those of the orbit that has
disappeared. The same order on both sides of the gap is thus not the signature of a
symmetry-breaking bifurcation: the two orbits are not connected by any
branch. The complete sequence leading to the large chaotic attractor, which
is absent from the pitchfork scenario, unfolds within
$\Delta\gamma\approx5\times10^{-4}$, the step of the full sweep, and is
therefore invisible in Fig.~\ref{fig:bifurcacion_omega13}.

Because no symmetry is broken, $\eta$ is of no use as an order
parameter here, and the natural observable is the critical multiplier itself.
Once the two multipliers of the orbit are real
($\gamma\gtrsim 0.42364 $), the dominant multiplier follows from the Lyapunov exponent as
$\mu=\exp(\lambda_{\max}kT_d)$; Fig.~\ref{fig:fold} resolves the approach
to the critical point with a step $\Delta\gamma=10^{-6}$. The saddle-node
normal form predicts
$1-\mu\propto(\gamma_{\mathrm{SN}}-\gamma)^{1/2}$, so that
$\lambda_{\max}$ reaches zero with a vertical tangent
[Fig.~\ref{fig:fold}(a)], in contrast with the linear crossing of the
pitchfork [Fig.~\ref{fig:pitchfork}(c)]. Accordingly, $(1-\mu)^{2}$ vanishes
linearly in $\gamma$ [Fig.~\ref{fig:fold}(b)]; a fit of
$\gamma=\gamma_{\mathrm{SN}}-a(1-\mu)^{2}-b(1-\mu)^{3}$ over the points with
$\mu>0.85$ yields $\gamma_{\mathrm{SN}}=0.4266645(2)$; adding a quartic term
shifts this value by less than $10^{-7}$, the estimate is stable for fitting
windows between $\mu>0.7$ and $\mu>0.9$, and it coincides with an independent
Newton continuation of the periodic orbit through the fold
(Appendix~\ref{subsec:newton}). A
term linear in $1-\mu$, which would dominate for a pitchfork or a
transcritical bifurcation, is compatible with zero.
The logarithmic slope of
$1-\mu$ against $\gamma_{\mathrm{SN}}-\gamma$ [Fig.~\ref{fig:fold}(c)]
increases from $0.44$ for $\mu>0.6$ to $0.48$ for $\mu>0.85$, the value
expected over that window from the fitted normal form, whose cubic term is
not negligible: the correction to $\nu=1/2$ is proportional to $1-\mu$ and
therefore decays only as $(\gamma_{\mathrm{SN}}-\gamma)^{1/2}$. Unlike the
pitchfork, where the square-root law holds over two decades of
$\gamma-\gamma_{c}$, the asymptotic regime of the fold is confined to
$\gamma_{\mathrm{SN}}-\gamma\lesssim3\times10^{-5}$.

The behavior of the antiperiodicity residual completes the
contrast [Fig.~\ref{fig:fold}(d)]. The residual stays at its numerical floor,
$\eta\approx2.5\times10^{-11}$, up to the last point of the branch, with no
counterpart of the pre-critical ramp of Fig.~\ref{fig:pitchfork}(d): critical
slowing down is just as severe at the fold, but the slowly relaxing mode is
$S$-even and does not contribute to $\eta$. Beyond $\gamma_{\mathrm{SN}}$ no
branch connects to the annihilated orbit, and the residual changes
discontinuously. The trajectory first falls onto a chaotic attractor over the
narrow interval $0.426665\lesssim\gamma\lesssim0.426708$
[first shaded band in Fig.~\ref{fig:zoom_fold}];
at its upper end a conjugate pair of asymmetric orbits of the same order
$k=5$, with $\eta=0.79$, is created by a saddle-node bifurcation of its own,
unrelated to the symmetric branch. Being asymmetric, this pair is free to
period-double, and it does so from $\gamma\approx0.42691$,
leading to a narrow-band chaotic attractor at $\gamma\approx 0.42702$
that expands abruptly into the large attractor at
$\gamma\approx0.42713$.
The two ways out of antiperiodicity are therefore distinguished by their
signatures in both diagnostics: at a pitchfork $\lambda_{\max}$ crosses zero
linearly and $\eta$ grows continuously as $(\gamma-\gamma_{c})^{1/2}$ on a
branch of the same order, whereas at a fold $\lambda_{\max}$ vanishes as a
square root and $\eta$ jumps by more than ten orders of magnitude onto an
unrelated attractor. In neither case does the antiperiodic orbit
period-double, in agreement with Ref.~\cite{swift1984suppression}.

\section{Conclusions}

In this work we have characterized the origin and distribution of
antiperiodicity in the Duffing--Holmes oscillator, combining a symmetry
analysis of the equations of motion with an extensive numerical
exploration of phase and parameter space
(Secs.~\ref{sec:dynamics}--\ref{sec:fold}). The results
address the three questions posed in the Introduction as follows.

(i) \emph{Antiperiodicity and conjugate
orbits.}
The equations of motion are invariant under the
half-period shift symmetry $S$, and within the framework of Swift and
Wiesenfeld~\cite{swift1984suppression} this invariance leaves every
periodic solution of the Duffing oscillator only two options: either it
is itself $S$-invariant, and hence antiperiodic, or it breaks the
symmetry and occurs as a member of a pair of conjugate solutions
exchanged by $S$. This dichotomy is rooted in two independent
ingredients of the model: the restoring force is odd (equivalently, the
potential is even, $V(-x)=V(x)$) and the drive $\gamma\cos(\omega t)$ is
itself antiperiodic. The invariance by itself guarantees that $S$ maps
solutions onto solutions; it does not guarantee the existence of
antiperiodic solutions, still less their stability. Existence for this
class of odd equations forced by an antiperiodic term is established,
under additional hypotheses, by the theorems of
Haraux~\cite{haraux1989antiperiodic}, whereas the parameter regions in
which stable antiperiodic orbits are actually realized were determined
here numerically. Compatibility alone,
however, is not the whole story: it is the nonlinearity that turns
antiperiodicity into a nontrivial phenomenon, opening subharmonic
responses and the hierarchy of antiperiodic waveforms with multiple
peaks per antiperiod, whereas in the linear limit $\beta=0$ the only
antiperiodic response is the trivial one locked at the driving
frequency (unstable for $\alpha<0$; Sec.~\ref{sec:existence}).

(ii) \emph{Organization in parameter space and well occupancy.}
Antiperiodic orbits occupy extended, well-defined regions of the
$(\omega,\gamma)$ plane rather than isolated points, interleaved with
periodic windows and chaotic domains (Figs.~\ref{fig:shrimp_regime}
and~\ref{fig:origen_regime}), and every antiperiodic orbit found
explores both potential wells, as expected for a symmetry that
exchanges them. Their waveforms and periods obey, in addition, a strict
regularity: invariance under $S$ imposes a parity selection rule on the
locking ratio, $T_{\mathrm{orb}}=k\,T_d$ with $k$ odd, verified without
exception across all parameter sweeps (Fig.~\ref{fig:sombreada}). The
rule is necessary but not sufficient: odd-$k$ windows may instead host
periodic orbits that break the symmetry spontaneously, in which case
they always occur as conjugate pairs $B=S(A)$ that jointly restore the
invariance lost by each orbit individually (Fig.~\ref{fig:mecanismo}).

(iii) \emph{Termination of antiperiodicity.} Among the generic local codimension-one
bifurcations, and under the assumptions of Sec.~\ref{sec:fold}, only two
exits are open to an $S$-invariant orbit; global transitions and jumps
between coexisting attractors are not excluded by this local analysis.
Both local exits are realized in the system. In the first, the exit from the antiperiodic window analyzed in detail involves neither chaos
nor period doubling. Across the transition the orbit retains its order
$k$, while the antiperiodicity residual $\eta$, which plays the role of
an order parameter, grows from zero with the mean-field exponent,
$\eta\propto(\gamma-\gamma_{c})^{1/2}$, and the maximum Lyapunov
exponent vanishes at $\gamma_{c}$ with the slope ratio dictated by the
pitchfork normal form (Figs.~\ref{fig:zoom_omega13}
and~\ref{fig:pitchfork}). The antiperiodic orbit thus sheds its
symmetry through a supercritical pitchfork bifurcation, splitting into
a conjugate pair $A$, $B=S(A)$ of the same order
(Sec.~\ref{sec:transition}). This is a direct, quantitative
realization of the theorem of Swift and
Wiesenfeld~\cite{swift1984suppression} (Sec.~\ref{sec:existence}): period
doubling is preceded by the pitchfork and confined to the symmetry-broken
branch~\cite{novak1982transition,dhumieres1982chaotic}.
In the second exit, realized by the $k=5$ window at the same frequency,
the antiperiodic
orbit is annihilated in a saddle-node bifurcation with an antiperiodic saddle
at $\gamma_{\mathrm{SN}}$: no symmetry is broken, $\eta$ stays at
its numerical floor up to the bifurcation, the critical multiplier approaches
unity as $(\gamma_{\mathrm{SN}}-\gamma)^{1/2}$, and the system jumps onto an
unrelated attractor (Sec.~\ref{sec:fold}, Fig.~\ref{fig:fold}).

Quantitatively, antiperiodic orbits occupy $56\%$ of the grid of
Fig.~\ref{fig:global}, $11\%$ of that of
Figs.~\ref{fig:shrimps_maps}--\ref{fig:shrimp_regime}, $41\%$ of that of
Figs.~\ref{fig:origin_maps}--\ref{fig:origen_regime}, and $28\%$ of the
frequency sweep of Fig.~\ref{fig:bifurcacion_gamma1}; none of the
$5.8\times10^{5}$ antiperiodic grid points, nor of the $2\,811$
antiperiodic points of the sweep (orders $k=1$ to $45$), has even
order, and all of them visit both wells. The
symmetry-breaking transition analyzed at $\omega=1.3$ occurs at
$\gamma_{c}=0.629587$ with exponent
$\nu=0.501\pm0.002$, and the
saddle-node termination of the $k=5$ window at the same frequency at
$\gamma_{\mathrm{SN}}=0.4266645(2)$.

Antiperiodicity thus emerges not as an accidental property of
particular waveforms but as the orbit-level manifestation of a
discrete symmetry of the driven system, and is expected to appear
generically in systems sharing this symmetry, such as the modified Chua
circuit in which antiperiodic oscillations were first observed
experimentally~\cite{freire2013antiperiodic,freire2014self}. Future work
will address the bifurcation mechanisms responsible for the generation
of additional peaks per antiperiod, and the routes by which the
symmetry-broken orbits born from antiperiodic ones reach chaos.
The multistability of the system, i.e.\ the coexistence of
antiperiodic, periodic, and chaotic attractors for the same forcing
parameters, which is responsible for the jumps between branches observed
in the continuation sweeps of Figs.~\ref{fig:bifurcacion_omega13}
and~\ref{fig:bifurcacion_gamma1}, raises the complementary question of how
these attractors share the space of initial conditions, i.e.\ of the
geometry and the relative volumes of their basins of attraction. This
question, which requires a different numerical strategy (independent
integration of large samples of initial conditions instead of
continuation), will be addressed in a separate paper.

\begin{acknowledgments}
A.C.M.\ acknowledges financial support from Project CSIC I+D \textit{Predictabilidad, caos, regularidad y simetr\'ias en sistemas f\'isicos no lineales} (No.~22520240100022UD), funded by CSIC--UdelaR. E.D.L. acknowledges support from Brazilian agencies CNPq (No. 304398/2023-3) and FAPESP (No. 2021/09519-5, No. 2025/27957-0). The authors acknowledge computing time on the high-performance cluster \texttt{ClusterUY}.
\end{acknowledgments}

\appendix

\section{Numerical methods: integration, Lyapunov spectra, and orbit classification}
\label{app:numerical}

\subsection{Model and time integration}
\label{subsec:integration}

Equation~\eqref{eq:duffing}, with the canonical double-well parameters
fixed in Sec.~\ref{sec:dynamics} and the forcing amplitude $\gamma$ and
angular frequency $\omega$ as control parameters, is written as the
non-autonomous first-order system
\begin{equation}
  \dot{x} = y, \qquad
  \dot{y} = -\delta y - \alpha x - \beta x^3 + \gamma\cos(\omega t).
  \label{eq:system}
\end{equation}
All computations were carried out in Julia with the
\textsc{DynamicalSystems.jl} ecosystem~\cite{Datseris2018}, which provides a
unified interface to the high-order solvers of \texttt{OrdinaryDiffEq.jl}.
Trajectories are integrated with the ninth-order explicit Runge--Kutta scheme
of Verner (\texttt{Vern9}) under adaptive step-size control, with absolute and
relative tolerances both set to $10^{-9}$; the high order and tight tolerances
are required to resolve the long-time dynamics faithfully and to obtain
accurate Lyapunov estimates near the regular--chaotic boundary. For each
parameter pair $(\omega,\gamma)$ an initial transient of duration
$T_{\mathrm{tr}}=500$ is discarded so that the orbit settles onto its
attractor, and the asymptotic trajectory is then recorded over a window
$T_{\mathrm{int}}=1.4\times10^{4}$ sampled on a uniform output grid of spacing
$\Delta t=10^{-2}$.

\subsection{Lyapunov exponents and identification of chaos}
\label{subsec:lyapunov}

The chaotic or regular nature of each orbit is established from the Lyapunov
spectrum, obtained by integrating the variational (tangent) equations of
Eq.~\eqref{eq:system} together with a set of perturbation vectors that are
periodically re-orthonormalized (Benettin's algorithm), accumulated over a
time $T_{\lambda}=10^{4}$ after the transient. The tangent-space formulation
is preferred over the two-trajectory separation method because the latter
becomes unreliable when the perturbation size approaches the integration
tolerance. Viewed as a dissipative two-dimensional flow, the oscillator
possesses two nontrivial exponents $\lambda_{1}\ge\lambda_{2}$ (the exponent
associated with the time direction of the extended autonomous system vanishes
identically and is omitted), whose sum is fixed by the constant phase-space
contraction rate of the flow,
\begin{equation}
  \lambda_{1}+\lambda_{2}
  = \big\langle \nabla\!\cdot\mathbf{f} \big\rangle
  = -\delta ,
  \label{eq:trace}
\end{equation}
with $\mathbf{f}$ the vector field of Eq.~\eqref{eq:system},
independently of the forcing parameters. This exact relation, recovered by the
simulations to within the integration tolerance, provides a stringent
consistency check on the numerics and yields the second exponent at no
additional cost as $\lambda_{2}=-\delta-\lambda_{1}$.

An orbit is classified as chaotic whenever
$\lambda_{\max}=\lambda_{1}>\varepsilon_{\lambda}$, with the small positive
threshold $\varepsilon_{\lambda}=10^{-3}$ chosen empirically from the
distribution of computed exponents so as to separate the chaotic band from the
numerically vanishing values that characterize periodic motion. Orbits below
the threshold are forwarded to the periodicity analysis described next.

\subsection{Stroboscopic section and order of the orbit}
\label{subsec:strobo}

Regular orbits are characterized through the stroboscopic Poincar\'e map,
built by sampling the state at integer multiples of the forcing period
$T_d$,
\begin{equation}
  \mathbf{P}_{n} = \bigl(x(t_{n}),\,y(t_{n})\bigr), \qquad t_{n}=nT_d .
\end{equation}
Because the strobe instants do not in general coincide with grid points, the
state is evaluated there by cubic Hermite interpolation, using the exact
derivative $\dot{x}=y$ as the interpolation slope; this yields fourth-order
accuracy and removes the sampling jitter that would otherwise limit the
attainable tolerance.

A period-$k$ orbit visits $k$ distinct points of the section, so that
$\mathbf{P}_{n}\approx\mathbf{P}_{n+k}$ for all $n$ in the asymptotic regime,
and its period is $T_{\mathrm{orb}}=kT_d$, where the integer $k$ is the
order of the orbit defined in Eq.~\eqref{eq:order}. The order is determined in two steps. First,
a candidate lag $g$ is identified as the smallest one for which
$\mathbf{P}_{1+g}$ returns to within a tolerance $\varepsilon$ of the
reference point $\mathbf{P}_{1}$ (first return). Second, the candidate is
accepted only if the recurrence holds \emph{globally}, i.e.\ if the
$90^{\text{th}}$ percentile of the distances
$\{\lVert\mathbf{P}_{n}-\mathbf{P}_{n+g}\rVert$ falls below $\varepsilon$;
the use of a high percentile rather than the maximum tolerates residual
transients while rejecting non-recurrent sequences. The tolerance is referred
to the size of the attractor,
\begin{equation}
  \varepsilon = \varepsilon_{r}\,L, \qquad
  L = \sqrt{(\Delta x)^{2}+(\Delta y)^{2}},
\end{equation}
with $\Delta x$, $\Delta y$ the phase-space extents of the recorded orbit and
$\varepsilon_{r}=10^{-2}$. Normalizing by the orbit size $L$, rather than by
the spread of the section itself, is essential: for a period-one orbit the
section collapses onto a single point and a tolerance referred to its
vanishing spread would yield spurious large periods. At least three
repetitions of the candidate cycle are required within the asymptotic window
for an orbit to be accepted as periodic (in practice, the
candidate lag is searched only up to $g\le N_{P}/3$, with $N_{P}$ the number
of stroboscopic samples retained after the transient, so that the accepted
cycle fits at least three times within the record); sequences that satisfy neither the
recurrence test nor the chaotic criterion are left unresolved.

Such unresolved points are shown in dark gray in the two-parameter
maps. They are rare: $0.085\%$ of the grid in Fig.~\ref{fig:global},
$0.43\%$ in Figs.~\ref{fig:shrimps_maps}--\ref{fig:shrimp_regime}, and
$0.14\%$ in Figs.~\ref{fig:origin_maps}--\ref{fig:origen_regime}. Between
a third and a half of them ($38\%$, $35\%$, and $48\%$, respectively) have
$\lambda_{\max}>-0.01$, i.e.\ they lie at the edge of
the chaos threshold; the remainder correspond to long transients that
occupy a sizeable fraction of the recorded window after a jump between
coexisting attractors along the continuation path, so that the recurrence
test fails although the final state is regular.

\subsection{Waveform descriptors: peaks, antiperiodicity and well occupancy}
\label{subsec:descriptors}

For each periodic orbit three further descriptors are recorded on the
late-time portion of the trajectory. The number of peaks $n_{\mathrm{pk}}$ is
the count of local maxima of $x(t)$ within one response period
$T_{\mathrm{orb}}$, a peak being identified as a sample larger than both of
its neighbours; together with the order $k$, this count distinguishes the
different periodic families found in the $(\omega,\gamma)$ plane.

Antiperiodicity is
tested through the half-period condition of
Sec.~\ref{sec:dynamics}, which automatically
guarantees $T_{\mathrm{orb}}$-periodicity. It is quantified by the
amplitude-normalized residual $\eta$ of Eq.~\eqref{eq:antip_residual},
with the averaging window and the normalization defined there; the
orbit is flagged as antiperiodic whenever $\eta<\varepsilon_{a}$, with
$\varepsilon_{a}=10^{-2}$ again fixed from the simulations.

Finally, the well occupancy is determined from the sign of $x$ over the final
window of the trajectory, of duration
$T_{\mathrm{w}}=\max(100,\,T_{\mathrm{orb}})$ time units for periodic orbits
and $T_{\mathrm{w}}=100$ for chaotic ones, so that the window always spans at
least one full period of a periodic orbit: orbits confined to a single well keep a definite
sign of $x$, whereas cross-well orbits change sign.

\subsection{Parameter-plane scan with attractor continuation}
\label{subsec:scan}

The $(\omega,\gamma)$ plane is mapped on a dense rectangular grid, with
$\omega$ varied in an outer loop and $\gamma$ in an inner loop. Within each
value of $\omega$, the amplitude is increased adiabatically: the initial
condition for each $\gamma$ is taken to be the final state reached at the
previous $\gamma$, so that the computation tracks a single attractor branch by
numerical continuation and minimizes jumps between coexisting
attractors. Whenever $\omega$ is changed, the initial condition is reset to a
fixed reference state $\mathbf{u}_{0}=(x_0,\dot{x}_0)=(2.0,\,0.1)$. common to all the continuation sweeps of this work. Because the continuation makes the
inner ($\gamma$) loop intrinsically sequential while the outer ($\omega$) loop
is independent, the scan parallelizes trivially over $\omega$ across threads,
each thread carrying out the full $\gamma$ continuation for its assigned
frequencies.

In all the continuation sweeps of this work the continuation is carried out
coherently with the forcing: the transient and the total integration time
spent at each parameter value are rounded up to integer numbers of forcing
periods $T_d$, so that the state handed over to the next parameter value,
and used to seed the separate integration that yields $\lambda_{\max}$,
corresponds to the same forcing phase $\phi_0$ at which the stroboscopic
section is taken. Without this precaution the phase mismatch acts as a
finite kick at every parameter step; wherever a chaotic saddle coexists
with the attractor, such kicks trigger long chaotic transients that bias
$\lambda_{\max}$ and may divert the continuation to a different
attractor; in the frequency sweep of Fig.~\ref{fig:bifurcacion_gamma1}
they produced isolated spurious antiperiodic points that roughly tripled
the number of boundaries between antiperiodic and non-antiperiodic
points.

This continuation strategy is common to all the $(\omega,\gamma)$
sweeps reported in this work, one-parameter bifurcation diagrams and
two-parameter maps alike; the fixed-initial-condition scans used in the
robustness test of Appendix~\ref{subsec:robustness} are the only
computations in which neighboring grid points are integrated
independently.

\subsection{Robustness with respect to the sweep strategy}
\label{subsec:robustness}

To assess the robustness of the dynamical classification with respect to
the sweep strategy, we compared three independent scans of the
$(\omega,\gamma)$ plane obtained from the same initial condition
$\mathbf{u}_{0}$: attractor tracking along $\omega$ at fixed $\gamma$,
attractor tracking along $\gamma$ at fixed $\omega$, and fixed initial
conditions (the same $\mathbf{u}_{0}$ reintroduced at every grid point,
with no tracking). The aggregate statistics, namely the fractions of
points classified as chaotic, periodic, or antiperiodic and the
distribution of $\lambda_{\max}$, are essentially insensitive to the
sweep strategy. In marked contrast, the well selected by single-well
trajectories depends strongly on whether the attractor is tracked and
along which axis, the disagreement concentrating in the regions of the
plane where both wells are simultaneously accessible. We interpret this
as a direct signature of multistability-induced discontinuities:
tracking the attractor along a given axis imposes a
sweep-direction-dependent hysteresis on the basin selected at each
$(\omega,\gamma)$, so that a well map obtained from a single tracking
direction reflects sweep history rather than an intrinsic property of
the point; fixed initial conditions remove this hysteresis, but at the
cost of requiring the transient to reconverge to the attractor from
scratch at every grid point. Claims about well identity are therefore
restricted to regions where the three strategies agree.

\subsection{Newton continuation of the
antiperiodic orbit through the fold}
\label{subsec:newton}

The location of the saddle-node bifurcation of
Sec.~\ref{sec:fold} was checked independently of the continuation
sweeps, which cannot follow an unstable orbit. The antiperiodic orbit of
order $k=5$ was computed as a fixed point of the stroboscopic map,
$\mathbf{u}-P^{k}(\mathbf{u};\gamma)=\mathbf{0}$, by Newton's method in
an independent code (fixed-step fourth-order Runge--Kutta integration of
Eq.~\eqref{eq:system} and of its variational equations, which provide
the Jacobian $DP^{k}$ and the Floquet multipliers), and the branch was
followed in $\gamma$ by pseudo-arclength
continuation~\cite{seydel2010practical}, which passes regularly through
the fold and onto the unstable saddle branch. The dominant real
multiplier approaches $+1$ at the fold along the stable branch and
exceeds unity along the saddle branch. The $S$-invariance of the saddle
branch was established by evaluating the antiperiodicity residual $\eta$
of Eq.~\eqref{eq:antip_residual} on the orbit reconstructed from each
converged fixed point: it remains at the numerical floor on both
branches, i.e.\ the saddle annihilated at the fold is antiperiodic. The
turning point located in this way coincides, to within $10^{-7}$, with
the value $\gamma_{\mathrm{SN}}=0.4266645(2)$ obtained from the fit of
Sec.~\ref{sec:fold}.

\bibliography{/home/arturo/Dropbox/bibtex/mybib}

%
%

\end{document}